\documentclass[conference]{IEEEtran}
\usepackage{cite}
\usepackage{amsmath,amssymb,amsfonts}
\usepackage{algorithmic}
\usepackage{graphicx}
\usepackage{subcaption}
\usepackage{textcomp}
\usepackage{amsthm}
\newtheorem{theorem}{Theorem}
\usepackage{multirow}
\usepackage[table,xcdraw]{xcolor}
\usepackage{siunitx}
\usepackage{tablefootnote}
\usepackage{footmisc}
\usepackage{url}
\newcommand{\yy}{\mathbf{y}}
\newcommand{\GenRE}{\mathbf{GenRE}}
\newcommand{\xx}{\mathbf{x}}
\newcommand{\ww}{\mathbf{w}}
\newcommand{\ff}{\mathbf{f}}
\newcommand{\RR}{\mathbf{R}}
\newcommand{\DD}{\mathbf{D}}
\newcommand{\HH}{\mathbf{H}}
\newcommand{\BB}{\mathbf{B}}
\newcommand{\alp}{\boldsymbol\alpha}
\newcommand{\Spsi}{\mathbf{\psi}}
\newcommand{\Cpsi}{\mathbf{\Psi}}
\newcommand{\qq}{\mathbf{q}}
\newcommand{\g}{\mathbf{g}}
\newcommand{\h}{\mathbf{h}}
\newcommand{\gTild}{\mathbf{\widetilde{g}}}
\newcommand{\hTild}{\mathbf{\widetilde{h}}}
\newcommand{\gPri}{\mathbf{g'}}
\newcommand{\hPri}{\mathbf{h'}}
\newcommand{\gTildPri}{\mathbf{\widetilde{g'}}}
\newcommand{\hTildPri}{\mathbf{\widetilde{h'}}}
\newcommand{\GH}{\mathbf{GH}}
\newcommand{\HHm}{\mathbf{HH}}
\newcommand{\HG}{\mathbf{HG}}
\newcommand{\GG}{\mathbf{GG}}
\newcommand{\Cphi}{\mathbf{\Phi}}
\newcommand{\Lbtwo}{\frac{L}{2}}
\newcommand{\LbtwoX}{\frac{L}{2}\times512}

\newcommand{\I}{\mathbf{I}}
\newcommand{\QQ}{\mathbf{Q}}

\begin{document}	
	\title{Image Denoising in FPGA using Generic Risk Estimation\\}	
	\author{\IEEEauthorblockN{Rinson Varghese}
\IEEEauthorblockA{College of Engineering and Management, Punnapra\\
Kerala, India\\
Email: rinsonvarghese@cempunnapra.org}
\and
\IEEEauthorblockN{Chandrasekhar Seelamantula}
\IEEEauthorblockA{IISc, Bangalore\\
India\\
Email: css@iisc.ac.in}
\and
\IEEEauthorblockN{Rathna G N}
\IEEEauthorblockA{IISc, Bangalore\\
India\\
Email:  rathna@iisc.ac.in}
\and
\IEEEauthorblockN{Ashutosh Gupta}
\IEEEauthorblockA{SAC, ISRO\\
India\\
Email: ashutoshg@sac.isro.gov.in}
\and
\IEEEauthorblockN{Debajyoti Dhar}
\IEEEauthorblockA{SAC, ISRO\\
India\\
Email:debajyotid@isro.gov.in}}
	
\maketitle
	
\begin{abstract}
	The generic risk estimator addresses the problem of denoising images corrupted by additive white noise without placing any restriction on the statistical distribution of the noise. In this paper, we discuss an efficient FPGA implementation of this algorithm. We use the undecimated Haar wavelet transform with shrinkage parameters for each sub-band as the denoising function. The computational complexity and memory requirement of the algorithm  is first analyzed. To optimize the performance, a combination of convolution and recursion is employed to realize Haar filter bank and gradient descent algorithm is used to find the shrinkage parameters. A fully pipelined and parallel architecture is developed to achieve high throughput. The proposed design achieves an execution time of 3.5ms for an image of size $512\times512$. We also show that the recursive implementation of Haar wavelet is more expensive than the direct implementation in terms of hardware utilization.		
\end{abstract}	
\section{Introduction}
Images and videos are omnipresent in our lives. We see them on television, surveillance systems, medical imaging systems etc. In every image processing application, we extract some information from one or more images taken by the camera and process it further. Unfortunately, the fundamental physical laws which govern the conversion of light into pixel values in Charge Coupled Device (CCD) sensors introduce noise into the image~\cite{QSI}. There are several potential noise sources in a CCD camera, such as shot noise, dark current, pixel non-uniformity and quantization noise. In very low light applications, when there is a time constraint on exposure time, CCD senors are seriously degraded by their readout noise~\cite{wilkinson1998digital}. Hence the level of noise depends not only on the sensor, but on the ambient conditions as well. Estimation and removal of noise is thus essential to increase the performance of image processing algorithms in the subsequent stages.

Since noise is random in nature, it is characterized in the statistical setting described above by its distribution function. Additive White Gaussian noise (AWGN) is the most commonly used model in image processing. In fact, the dark current, which is dependent on the temperature and exposure time is modeled as a Gaussian distribution. Noise may follow other distributions as well, for instance, Shot and Read noise caused by on chip amplifiers follow Poisson distribution where as hot pixels follow impulsive distribution~\cite{healey1994radiometric}. Noise in magnetic resonance imaging is found to be Rician distributed~\cite{gravel2004method} and in SAR images it is speckle noise~\cite{frost1982model}. Hence there is a need for a denoising method which is independent of the noise distribution. Further, to use these algorithms in real time applications, we may have to meet some timing and power constraints. Hence we go for hardware implementation of the algorithm instead of software. FPGA is a good choice in this context since we can employ pipelining and parallelism to improve the speed at a reduced power. 

There are different works which show the application of FPGA in image denoising and image processing. In~\cite{SimpleFPGA}, some simple filters like median filter, smoothing filter and edge detection filter are implemented in FPGA. Different architectures of a bilateral filter are discussed in ~\cite{Bilateral1, Bilateral2, Bilateralreconfigurable}, where~\cite{Bilateralreconfigurable} is the first scalable FPGA implementation of the bilateral filter. A reconfigurable gaussian noise detection in images is proposed in~\cite{hegde2015adaptive}, where the real time dynamic reconfiguration of the algorithm is achieved through the processor core. In~\cite{di2013aidi}, an adaptive image denoising IP core based on FPGA is proposed where adaptive gaussian filter is employed for noise elimination. The denoising framework in most of these algorithms puts some restriction on the noise distribution. With the popularity of the wavelet transform for the past three decades, several algorithms have been developed in wavelet domain. Only a few of these algorithms have been implemented in FPGA. A hardware architecture for wavelet domain video denoising system is proposed in \cite{katona2005fpga} where the denoising is performed in wavelet domain using spatially adaptive bayesian shrinkage. 

In this work we develop a hardware architecture for generic risk estimator (GenRE) proposed in~\cite{subramanian2016distribution}, where there is no restriction assumed on the noise distribution. We use undecimated Haar wavelet transform together with shrinkage parameters as the denoising function.
\begin{figure*}
	\centerline{\includegraphics[width=1\textwidth]{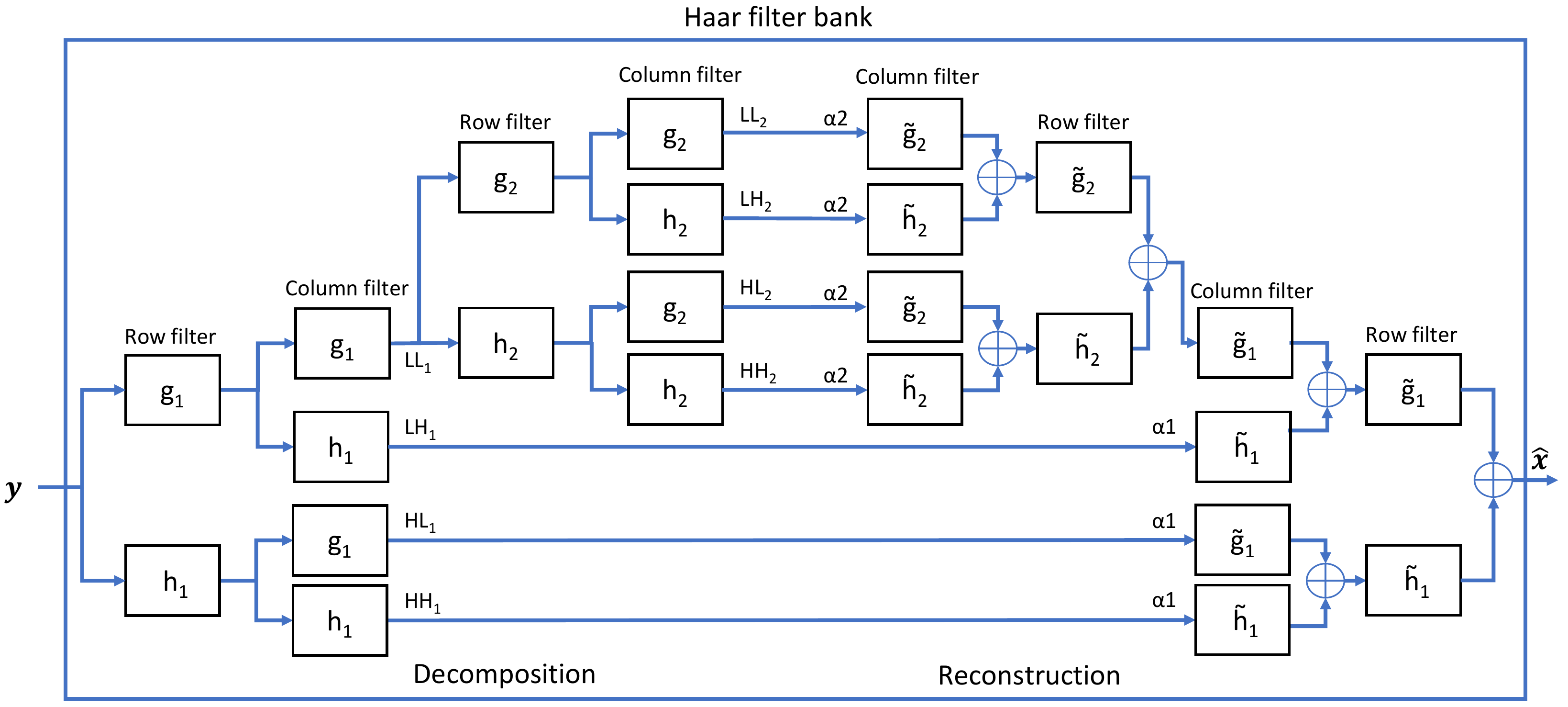}}
	\caption{A block diagram of an undecimated wavelet based denoising for two levels of decomposition with shrinkage parameters $\alpha_i$.}
	\label{fig-Haar_2_level}
\end{figure*}
\section{Generic Risk Estimator}
The objective of any image denoising algorithm is to estimate the unknown true image from a noisy observation. We model this problem as follows.
\begin{equation}
\yy = \xx + \ww
\end{equation}
where,\\
$\xx\in \mathbb{R^N}$ is the column-vectorized noise free image, \\ 
$\ww\in \mathbb{R^N}$ is the noise, whose first and second order statistics is known and assumed to have zero mean and variance.\\
$\yy\in \mathbb{R^N}$ is the noisy observation.\\
We also assume that $\ww$ is independent of $\xx$ and  the distribution of $\ww$ is unknown. The goal is to find an optimal denoising function $\ff: \mathbb{R^N}\mapsto \mathbb{R^N}$, which gives $\hat\xx$, an estimate of the true image , when operated on $\yy$. The optimality of the denoising function is determined by mean square error (MSE). Hence the whole problem can be restated as follows: Find a denoising function $\ff$ such that the MSE
\begin{equation}
\mathbf{C}(\ff) = \frac{1}{N}\mathbb{E}(||\xx - \hat{\xx}||^2)
\end{equation}
between $\xx$ and $\hat{\xx} = \ff(\yy)$ is minimized. However, non-availability of ground truth becomes an obstacle for direct calculation of the MSE. Hence instead of minimizing MSE, we find an unbiased estimate of the MSE and minimize it to find the denoising function.
\begin{theorem}
	Let $\ff$ be a linear denoising function, given as $\ff(\yy) = \BB\yy$, $\BB\in\mathbb{R}^{N{\times}N}$, then 	
	\begin{equation}
	\text{$\GenRE$}(\ff) = \frac{1}{N}(||\ff(\yy)-\yy||^2 + 2\sigma^2\text{Trace}(\BB)-N\sigma^2)
	\label{eq_Gen_RE_Th}
	\end{equation}
	is an unbiased estimator of the MSE.
\end{theorem}
A formal proof of the theorem is given in~\cite{subramanian2016distribution}.
\begin{figure*}
	\centerline{\includegraphics[width=1\textwidth]{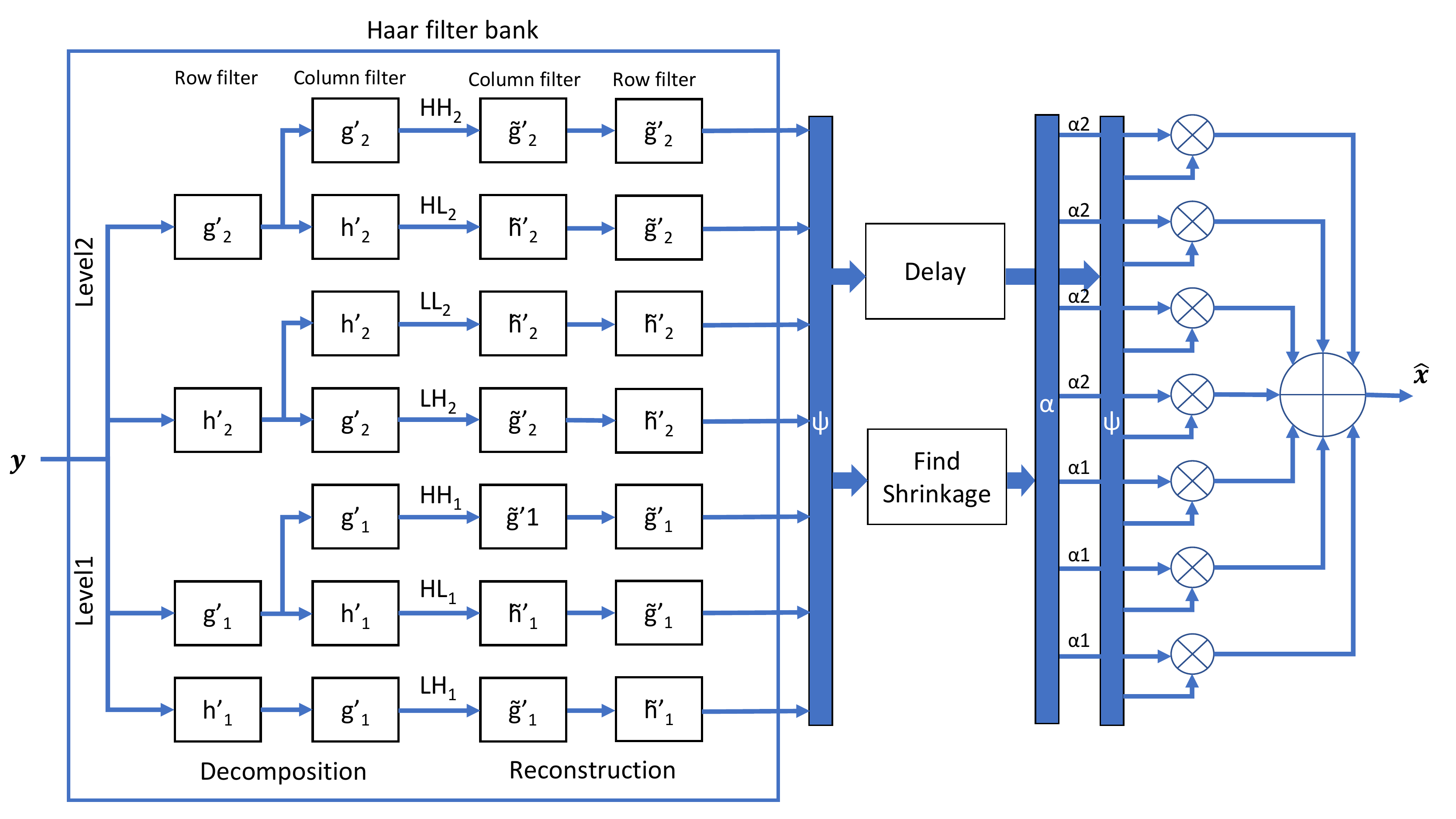}}
	\caption{A block diagram of GenRE based denoising using UWT.}
	\label{fig-Gen_RE_2_Level}
\end{figure*}
\section{GenRE based denoising using undecimated wavelet transform}
In discrete wavelet transform (DWT), at every level of decomposition we down-sample the signal which removes redundant information. Similar to this we up-sample the signal during reconstruction which restores the original sampling rate. Undecimated wavelet transform (UWT) removes this sampling rate conversion and instead of down-sampling the signal, the filter itself is up-sampled. The filter coefficients are upsampled by a factor of $2^{j-1}$ in the $j^{\text{th}}$ level of the algorithm. Fig.~\ref{fig-Haar_2_level} illustrates an undecimated Haar wavelet based denoising which uses two levels of decomposition. The denoising function is a combination of analysis filter bank, transform-domain shrinkage, and synthesis filter bank. In matrix form we represent the analysis filter bank as $\DD = [\DD_1^{\text{T}}, \DD_2^{\text{T}},\dots \DD_j^{\text{T}}]^{\text{T}}$ and synthesis filter bank as $\RR = [\RR_{\text{1}}, \RR_\text{2}, \dots \RR_\text{j}]$, such that $\RR\DD = \mathbf{I}$. Here both $\RR_\text{i}$ and $\DD_\text{i}$ are circulant matrices. We can describe the overall transform domain process as follows
\begin{eqnarray}
\nonumber
\hat{\xx} &=& \ff(\yy) = \sum_{i=1}^{j}\RR_i\alp_i\DD_i\yy = \sum_{i=1}^{j}\alp_i\HH_i\yy = \BB\yy\\
\hspace{0.25cm} &=& \sum_{i=1}^{j}\alp_i\Spsi_i = \Cpsi \alp 	
\end{eqnarray}
where $\HH_i = \RR_i\DD_i$, $\alpha_i$ is the shrinkage parameter for $\text{i}^{\text{th}}$ sub-band and $\{\Spsi_i\}_{i=1}^{J+1}$ forms the columns of $\Cpsi$. Here 
\begin{equation}
\BB = \sum_{\text{i = 1}}^{\text{j}}\alpha_i\HH_i = \sum_{\text{i = 1}}^{\text{j}}\alpha_i\RR_i\DD_i
\end{equation}
$\BB$ is a circulant matrix, since it is a product of two circulant matrices. Now we can rewrite \eqref{eq_Gen_RE_Th} as 
\begin{eqnarray}
\nonumber
\text{$\GenRE$}(\ff) &=& \frac{1}{N}(||\ff(\yy)-\yy||^2 + 2\sigma^2\text{Trace}(\BB)-N\sigma^2)\\	
\nonumber		
&=& \frac{1}{N}(||\Cpsi\alp-\yy||^2 + 2\sigma^2N\BB_{11}-N\sigma^2)\\
\nonumber
&=& \frac{1}{N}(||\Cpsi\alp-\yy||^2 + 2\sum_{\text{i = 1}}^{\text{j}}\alpha_i\HH_{11}^iN\sigma^2-N\sigma^2)\\	
&=& \frac{1}{N}(||\Cpsi\alp-\yy||^2 + 2\alp^\text{T}\qq-N\sigma^2)
\end{eqnarray}
We can find the shrinkage parameter $\alp$ by minimizing $\GenRE$, which gives
\begin{equation}
\nabla_\alpha(\text{$\GenRE$}(\ff)) = 2\Cpsi^T\Cpsi\alp^*-2\alp^\text{T}\yy-2\qq = 0	  
\end{equation}
and 
\begin{equation}
\alp^* = (\Cpsi^T\Cpsi)^{-1}(\Cpsi^\text{T}\yy - \qq)	
\label{eq-alpha}  
\end{equation}
Once $\alp$ is found, the denoised image can be estimated as
\begin{equation}
\hat{\xx} = \Cpsi\alp
\label{eq-xCap}  
\end{equation} 
A block diagram representation of the complete algorithm is shown in Fig.~\ref{fig-Gen_RE_2_Level}. Note that in the figure we have assumed a two level UWT for simplicity. For actual implementation we consider five levels of wavelet decomposition on images of size $512\times512$ which is row vectorized. 

We have two fundamental operations in the GenRE algorithm: Wavelet analysis of the image using undecimated Haar filter banks, which results in the matrix $\Cpsi$ and estimation of shrinkage parameter vector $\alp$. Since this involves compute-intensive operations like convolution, matrix inversion and multiplication, we should analyze them separately and use every possible means to reduce computational complexity. A complete analysis of this is done in the following sections.
\section{Computational complexity and Memory usage of Haar Filter Bank} 
Two approaches to find the wavelet sub-bands are shown in Fig.~\ref{fig-Haar_2_level} and Fig.~\ref{fig-Gen_RE_2_Level}. They are in fact similar except that the filters in the later one are formed by combining filters in a particular branch. For example, the first level Haar filters are  
\begin{alignat}{3}
\nonumber
\h_1 &= \frac{1}{2}\begin{bmatrix}
1, 1
\end{bmatrix},\quad
\g_1 &&= \frac{1}{2}\begin{bmatrix}
-1, 1
\end{bmatrix},\\
\hTild_1 &= \frac{1}{2}\begin{bmatrix}
1, 1
\end{bmatrix},\quad
\label{eq_Haar_coeff_lev1}
\gTild_1 &&= \frac{1}{2}\begin{bmatrix}
1, -1
\end{bmatrix}.
\end{alignat}
where $\h$ and $\hTild$ are the refinement filters in decomposition and recomposition phase, respectively. The corresponding wavelet filters are $\g$ and $\gTild$. Note that we have normalized the filters by its length. Since we use UWT, filter for other levels are found by up-sampling first level filters by $2^{j-1}$, where $j$ represents the level of filtering. Hence the filters for the second level are
\begin{alignat}{3}
\nonumber
\h_2 &= \frac{1}{2}\begin{bmatrix}
1, 0, 1, 0
\end{bmatrix},\quad 
&&\g_2 = \frac{1}{2}\begin{bmatrix}
-1, 0, 1, 0
\end{bmatrix},\\
\hTild_2 &= \frac{1}{2}\begin{bmatrix}
1, 0, 1, 0
\end{bmatrix},\quad 		
&&\gTild_2 = \frac{1}{2}\begin{bmatrix}
1, 0, -1, 0
\end{bmatrix}.
\end{alignat}
By using associativity of convolution and interchangeability of row and column operations, we can combine the filters in each branch as
\begin{align}
\left.\begin{aligned}
\hPri_1 &= \h_1 = \frac{1}{2}\begin{bmatrix}
1, 1
\end{bmatrix},\\
\hPri_2 &= \h_1*\h_2 = \frac{1}{4}\begin{bmatrix}
1, 1, 1, 1
\end{bmatrix},\\
\gPri_1& = \g_1 = \frac{1}{2}\begin{bmatrix}
-1, 1
\end{bmatrix},\\	
\gPri_2 &= \h_1*\g_2 = \frac{1}{4}\begin{bmatrix}
-1, -1, 1, 1
\end{bmatrix},\\	
\hTildPri_1& = \hTild_1 = \frac{1}{2}\begin{bmatrix}
1, 1
\end{bmatrix},\\
\hTildPri_2& = \hTild_1*\hTild_2 = \frac{1}{4}\begin{bmatrix}
1, 1, 1, 1
\end{bmatrix},\\
\gTildPri_1& = \gTild_1 = \frac{1}{2}\begin{bmatrix}
1, -1
\end{bmatrix},\\	
\label{eq_Haar_Comb_coeff_Decompose}
\gTildPri_2 &= \hTild_1*\gTild_2= \frac{1}{4}\begin{bmatrix}
1, 1, -1, -1
\end{bmatrix},
\end{aligned}\right\}
\end{align}
where $*$ represents convolution. Filters for the other levels follow the same structure but different lengths. The Haar filter bank in Fig.~\ref{fig-Gen_RE_2_Level} uses these combined filters. 

So far, the filters we discussed are all in 1D domain. We can also do the filtering in 2D domain, where the kernels are formed as outer products of 1D filters. The corresponding individual kernels are 
\begin{align}
\left.\begin{aligned}
\HG_1 &= \frac{1}{4}
\begin{bmatrix}
-1&-1\\
\phantom{-}1&\phantom{-}1
\end{bmatrix},\\
\GH_1 &= \frac{1}{4}
\begin{bmatrix}
-1&\phantom{-}1\\
-1&\phantom{-}1
\end{bmatrix},\\
\GG_1 &= \frac{1}{4}
\begin{bmatrix}
\phantom{-}1&-1\\
-1&\phantom{-}1
\end{bmatrix},\\
\HHm_1 &= \frac{1}{4}
\begin{bmatrix}
1&\phantom{-}1\\
1&\phantom{-}1
\end{bmatrix},\\
\HG_2 &= \frac{1}{4}
\begin{bmatrix}
-1&0&-1&0\\
\phantom{-}0&0&\phantom{-}0&0\\
\phantom{-}1&0&\phantom{-}1&0\\
\phantom{-}0&0&\phantom{-}0&0\\
\end{bmatrix},\\
\GH_2 &= \frac{1}{4}
\begin{bmatrix}
-1&0&1&0\\
\phantom{-}0&0&0&0\\
-1&0&1&0\\
\phantom{-}0&0&0&0\\
\end{bmatrix},\\
\GG_2 &= \frac{1}{4}
\begin{bmatrix}
\phantom{-}1&0&-1&0\\
\phantom{-}0&0&\phantom{-}0&0\\
-1&0&\phantom{-}1&0\\
\phantom{-}0&0&\phantom{-}0&0\\
\end{bmatrix},\\
\HHm_2 &= \frac{1}{4}
\begin{bmatrix}
1&0&1&0\\
0&0&0&0\\
1&0&1&0\\
0&0&0&0\\
\end{bmatrix},\\
\label{eq_Haar_coeff_2D_Decompose}
\end{aligned}\right\}
\end{align} and the combined kernels are 
\begin{align}
\left.\begin{aligned}
\HG'_1 &= \HG_1 = \frac{1}{4}
\begin{bmatrix}
-1&-1\\
\phantom{-}1&\phantom{-}1
\end{bmatrix},\\
\GH'_1 &= \GH_1 = \frac{1}{4}
\begin{bmatrix}
-1&\phantom{-}1\\
-1&\phantom{-}1
\end{bmatrix},\\
\GG'_1 &= \GG_1 = \frac{1}{4}
\begin{bmatrix}
\phantom{-}1&-1\\
-1&\phantom{-}1
\end{bmatrix},\\
\HHm'_1 &= \HHm_1 = \frac{1}{4}
\begin{bmatrix}
1&\phantom{-}1\\ 
1&\phantom{-}1
\end{bmatrix},\\	
\HG'_2 &= \HHm_1*\GG_2 = \frac{1}{16}
\begin{bmatrix}
-1&-1&-1&-1\\
-1&-1&-1&-1\\
\phantom{-}1&\phantom{-}1&\phantom{-}1&\phantom{-}1\\
\phantom{-}1&\phantom{-}1&\phantom{-}1&\phantom{-}1\\
\end{bmatrix},\\
\GH'_2 &= \HHm_1*\GH_2 = \frac{1}{16}
\begin{bmatrix}
-1&-1&\phantom{-}1&\phantom{-}1\\
-1&-1&\phantom{-}1&\phantom{-}1\\
-1&-1&\phantom{-}1&\phantom{-}1\\
-1&-1&\phantom{-}1&\phantom{-}1\\
\end{bmatrix},\\
\GG'_2 &= \HHm_1*\GG_2 = \frac{1}{16}
\begin{bmatrix}
\phantom{-}1&\phantom{-}1&-1&-1\\
\phantom{-}1&\phantom{-}1&-1&-1\\
-1&-1&\phantom{-}1&\phantom{-}1\\
-1&-1&\phantom{-}1&\phantom{-}1\\
\end{bmatrix},\\
\HHm'_2 &= \HHm_1*\HHm_2 = \frac{1}{16}
\begin{bmatrix}
1&1&1&1\\
1&1&1&1\\
1&1&1&1\\
1&1&1&1\\
\end{bmatrix}.
\label{eq_Haar_Comb_coeff_2D_Decompose}
\end{aligned}\right\}
\end{align}
The corresponding recomposition kernels can be obtained by flipping  the kernels \eqref{eq_Haar_coeff_2D_Decompose} and \eqref{eq_Haar_Comb_coeff_2D_Decompose} both vertically and horizontally. Kernels for the higher levels can be calculated in a similar way and they hold the similar structure.

Now we have four sets of filters given by \eqref{eq_Haar_coeff_lev1} -- \eqref{eq_Haar_Comb_coeff_2D_Decompose} and each of them can be used to find UWT of the image independently. We need to select one set of filters or a combination of them after analyzing computational complexity and memory requirement. Across all the levels, individual filters have only 2 and 4 non-zero elements for 1D and 2D case, respectively. Here, we use a convolution based filtering approach for hardware implementation. Furthermore, the non-zero elements in both these filters are all unity. Hence we can neglect the multiplication involved in the convolution. Combined filters, on the other hand have all its elements as unity with a change of sign for half of its elements. The change in the sign have some structure which is favorable for recursive implementation and hence we can employ either convolution or recursion based approaches. There is an additional multiplication with normalization factor involved in both the cases which we neglect from computation analysis, as it is a multiplication with power of 2 which can be achieved by shift operation. A detailed analysis of the computations and memory usage is done in the following sections.
\subsection{Analysis of Individual Filters}
\label{sec_individual filters_Analysis}
The difference equations for 1D individual filters in decomposition can be written as
\begin{equation}
y(n) = x(n) + x(n-\frac{L}{2}), 
\label{eq-1D_LPF}
\end{equation}
and 
\begin{equation}
y(n) = x(n-\frac{L}{2}) - x(n), 
\label{eq-1D_HPF}
\end{equation}
for refinement and wavelet filters, respectively,
\begin{figure}[h]
	\centerline{\includegraphics[width=1\columnwidth]{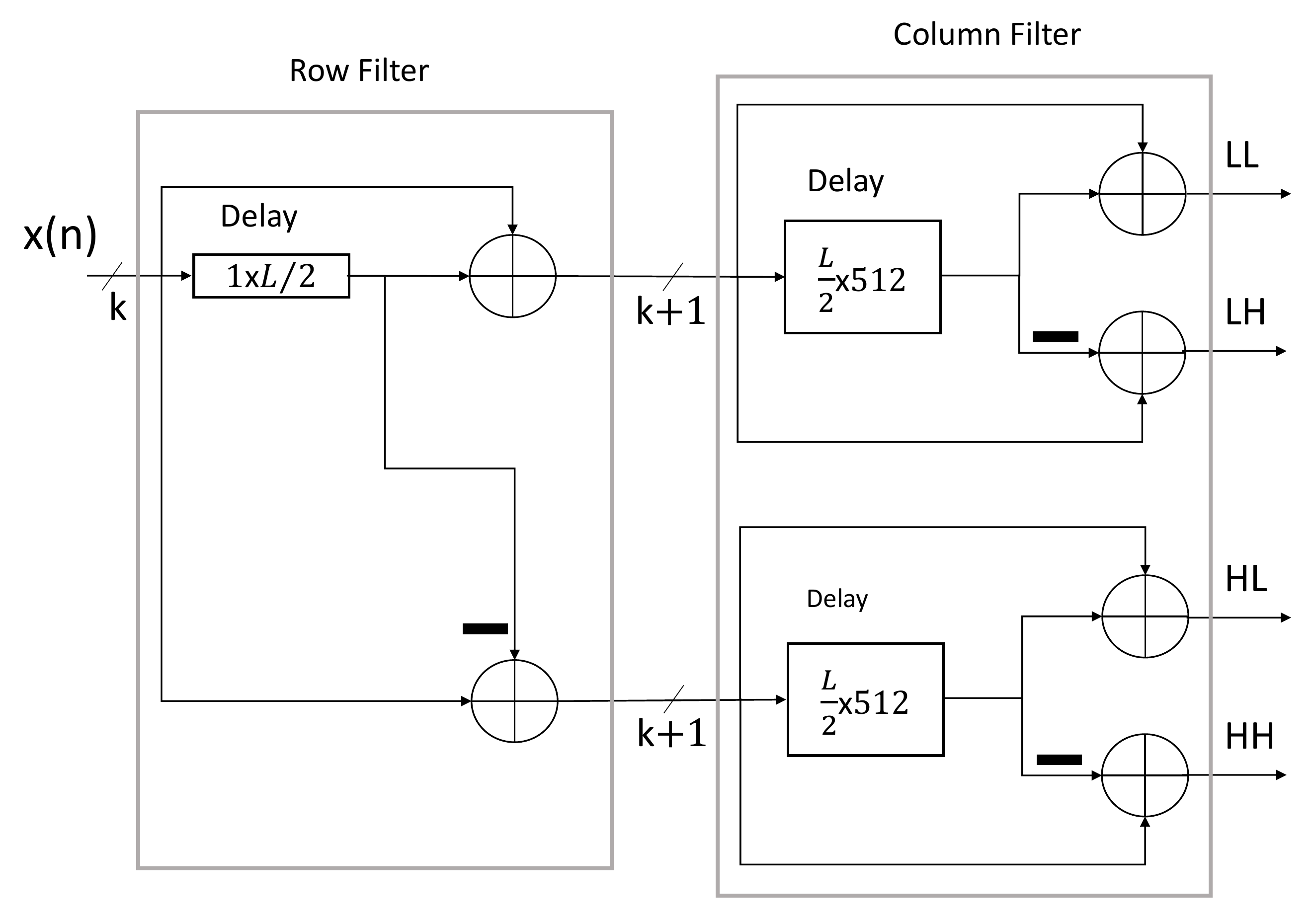}}
	\caption{Implementation of UWT using 1D individual filters. Here $x(n)$ represents row vectorized input image.}
	\label{fig-1D_UWT}
\end{figure}
where L is the length of the filter in the particular filtering level. Here we have one addition per output pixel. We need to apply the filter both on rows and columns, as depicted in Fig.~\ref{fig-1D_UWT}. Thus, there are a total of six additions involved in the calculation of the four sub-bands in a level. The average additions per output pixel is $6/4 = 1.5$. Further, the implementation of \eqref{eq-1D_LPF} and \eqref{eq-1D_HPF} requires delays of length $\frac{L}{2}$. Since the image is row-vectorized, we need $\frac{L}{2}\times512$ sized buffers at the input of the column filters to implement $\frac{L}{2}$ delay over columns. As a result, we need one $\Lbtwo$-length buffer and two $\LbtwoX$ buffers for the implementation of 1D filters in a particular level  as shown in Fig.~\ref{fig-1D_UWT}. 

\begin{figure}[h!]
	\centerline{\includegraphics[width=0.75\columnwidth]{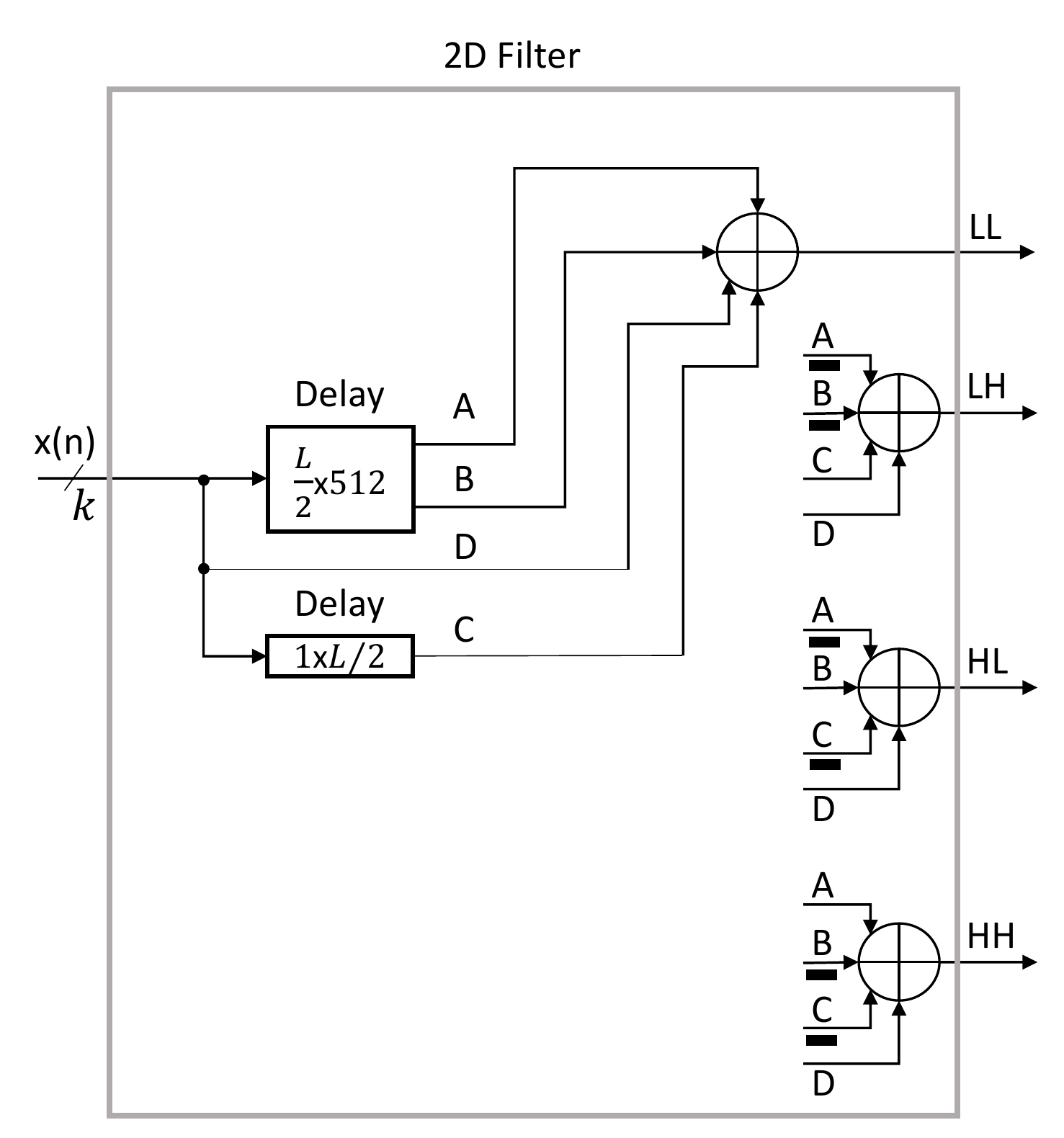}}
	\caption{Implementation of UWT using 2D individual filters. Here A, B, C and D represents $\I(n_1-\Lbtwo, n_2-\Lbtwo), \I(n_1, n_2-\Lbtwo), \I(n_1-\Lbtwo, n_2)$ and  
		$\I(n_1, n_2)$ respectively where $\I$ is the input image and $x(n)$ is the row vectorized image.}
	\label{fig-2D_UWT}
\end{figure}

We can do a similar analysis in 2D. Using \eqref{eq_Haar_coeff_2D_Decompose}, we write the equations for different sub-bands as 
\begin{align}
\left.\begin{aligned}
\text{LH}&=& \text{D}-\text{A}+\text{C}-\text{B},\\
\text{HL}&=& \text{D}-\text{A}-\text{C}+\text{B},\\
\text{HH}&=& \text{D}+\text{A}-\text{C}-\text{B},\\
\label{eq-2D}
\text{LL}&=& \text{D}+\text{A}+\text{C}+\text{B}.
\end{aligned}\right\}
\end{align}
Here A, B, C and D represent $\I(n_1-\Lbtwo, n_2-\Lbtwo), \I(n_1, n_2-\Lbtwo), \I(n_1-\Lbtwo, n_2)$ and  
$\I(n_1, n_2)$, respectively, where $\I$ is the input image. Fig.~\ref{fig-2D_UWT} illustrates the implementation of these equations. 
Here we can share the intermediate results D$-$A, D$+$A, C$-$B and C$+$B. Hence there are eight additions involved in the calculation of all the sub-bands and the average additions per output pixel is $\frac{8}{4} = 2$. Further, we need only two buffers here, one $\Lbtwo$ and another $\LbtwoX$ as compared to the 1D implementation. The bit-widths of these buffers will also be less by one.

The filters in recomposition side are similar except for a flip. As a result, the number of additions involved in filtering will also be the same. In contrast to decomposition, inputs to recomposition filters are the different sub-bands that we calculated in the decomposition phase. The results of these filters are then combined to give a single low pass band from a filtering level as shown in Fig.~\ref{fig-Haar_2_level}. Hence the average number of additions per output pixel will increase to 6 in 1D and 12 in 2D implementations. The buffer usage will also increase since we can not share the buffers as we did during decomposition. Here, we may have to use four $\Lbtwo$ and $\LbtwoX$ buffers for implementing the filters in a particular level. 
\subsection{Analysis of Combined Filters}
\begin{figure}
	\centerline{\includegraphics[width=1\columnwidth]{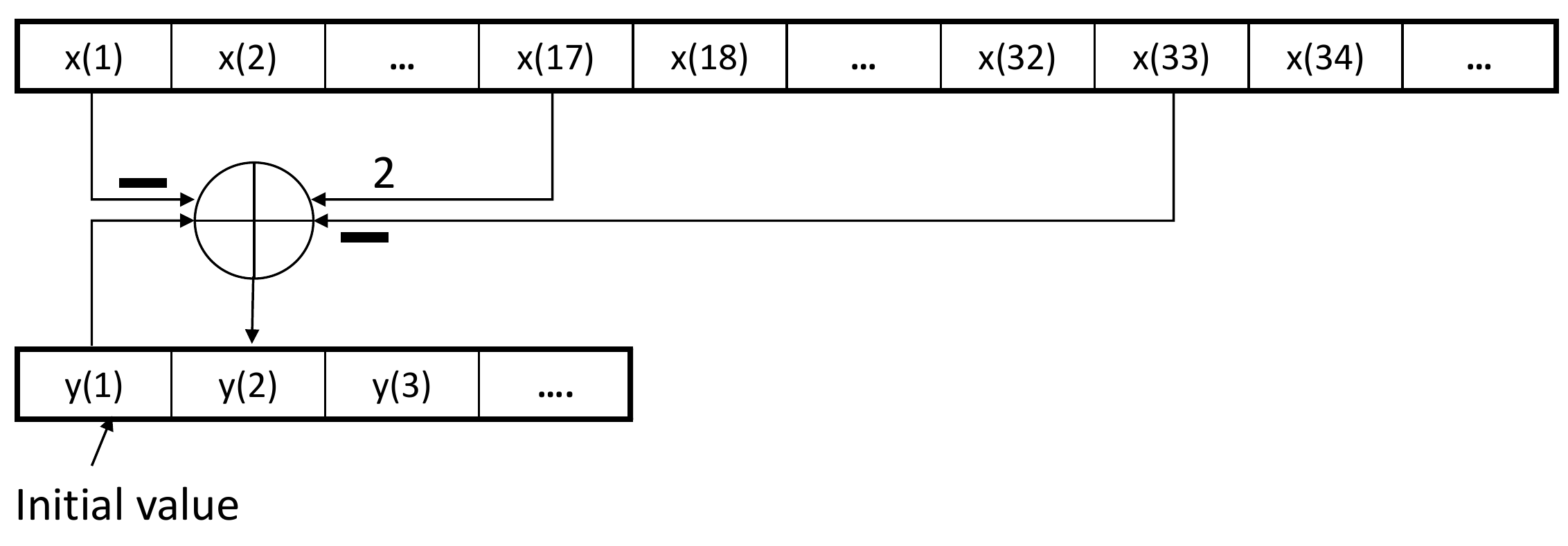}}
	\caption{Illustration of 1D recursive implementation of high pass filter in fifth level of Haar filter bank. The initial value is found using convolution.}
	\label{fig-1D_recursion_H_Filter_Lev5}
\end{figure}

\begin{figure*}
	\centerline{\includegraphics[width=1\textwidth]{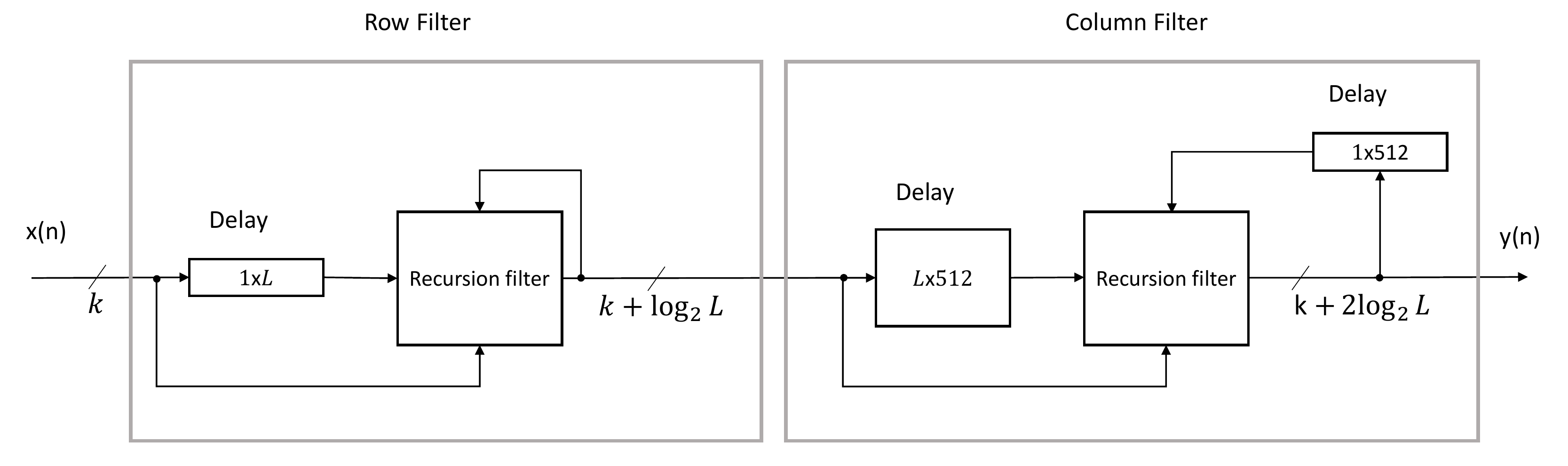}}
	\caption{Recursive implementation of filters in a single wavelet channel. The recursion filter could be either high pass or low pass depending on the channel.}
	\label{fig-1D_RUWT}
\end{figure*}

In contrast to the individual filters, combined filters have all elements with unity magnitude, which results in $L-1$ and $L^2-1$ additions in 1D and 2D domains, respectively. We can find the average additions per output pixel by doing a similar analysis as in section \ref{sec_individual filters_Analysis} with $L-1$ and $L^2 - 1$ replacing their counter parts in the respective domains. This will result in $\frac{6}{4}(L-1)$ and $L^2 - 1$ additions per output pixel respectively for 1D and 2D filters. As we have already mentioned, for these particular filters we can use recursion in place of convolution to reduce calculations. For instance, if we take the 1D wavelet filter in the fifth level, which is
\begin{equation}
\hPri_5 =\frac{1}{32}
\begin{bmatrix}
-1, -1, -1, \dots, 1, 1, 1, \dots
\end{bmatrix}_{1\times32}
\end{equation}
we can implement it recursively as depicted in Fig.~\ref{fig-1D_recursion_H_Filter_Lev5}. Here, the initial value is calculated using convolution and the remaining pixel values are found using recursion. The recursive equations for this implementation are
\begin{equation}
y(n) = y(n-1) - x(n-L) + x(n), 
\label{eq_1D_recursion_L_Filter}
\end{equation}
and
\begin{equation}
y(n) = y(n-1) - x(n-L) + 2x(n-\frac{L}{2}) - x(n),
\label{eq_1D_recursion_H_Filter}
\end{equation}
for refinement and wavelet filters, respectively, where L is the length of the filter. There are three additions in the wavelet filter and two in the corresponding refinement filter. Hence, calculation of LH and HL sub-bands require five additions each. To find LL and HH sub-bands we need only five more additions since we already have the partial results. The average additions per output pixel are $(15)/4 = 3.75$. Implementation of recursive filters for a single wavelet channel is shown in Fig.~\ref{fig-1D_RUWT}. Here, the requirement of delays is similar to the individual filter case except that instead of $\Lbtwo\times512$ delays we need $L\times512$ delays and we introduce a new delay of size $1\times512$ in the column filter side. Introduction of this new delay is caused by the row-vectorization of the input image and dependency of recursion on the previous result. Since the image is row-vectorized, we apply the column filter row-wise, i.e., we calculate the first result pixel in the first column and then the first pixel in the second column and so on. Once the first row is finished we start with the second row. Now in order to do a recursion, we need the result pixels in the previous row which is stored in the $1\times512$ buffer. Hence, altogether we need one $L$-length, two $L\times512$ and four $512$-length  buffers for the implementation of recursive filters in a particular level.

\begin{figure}[h]
	\begin{subfigure}[t]{.3\textwidth}
		\centerline{\includegraphics[width=0.8\columnwidth]{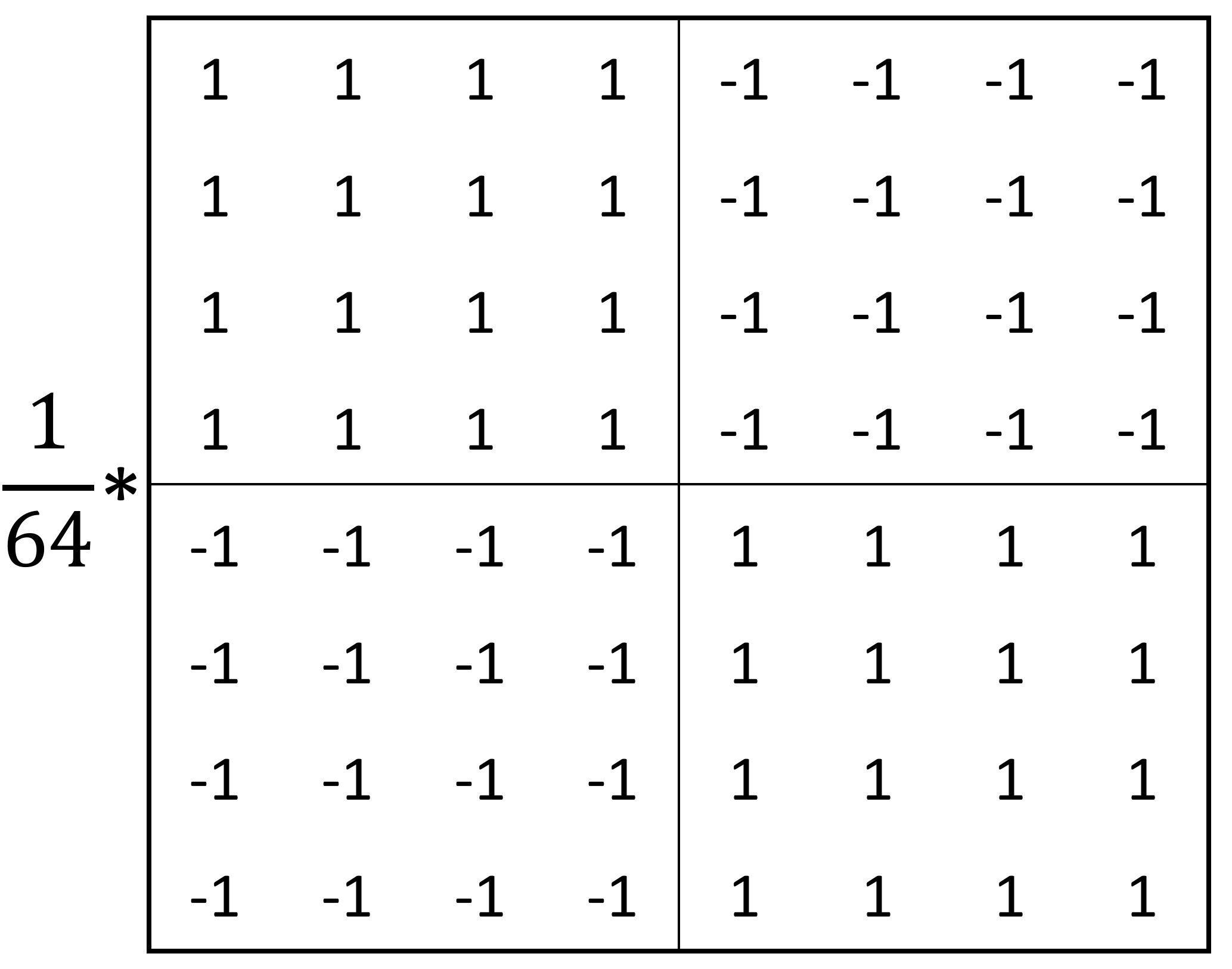}}
		\caption{2D kernel for HH sub-band in level 3.}
		\label{fig-2D_Kernel_HH_Lev3}
	\end{subfigure}%
	\begin{subfigure}[t]{.3\textwidth}
		\centerline{\includegraphics[width=0.7\columnwidth]{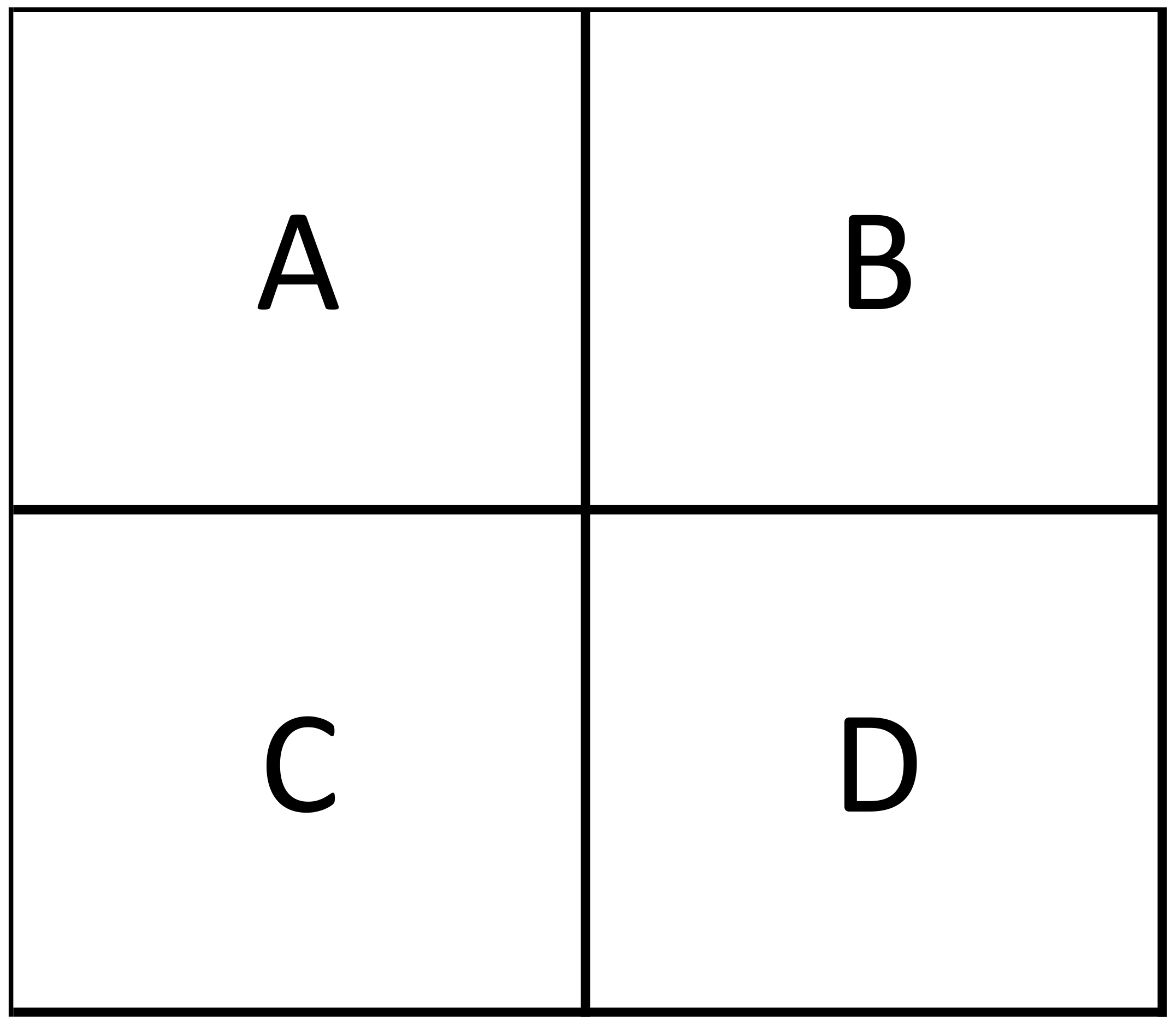}}
		\caption{Image pixels grouped corresponding to the filter structure.}
		\label{fig-Pixel_Grouping_Lev3}
	\end{subfigure}
	\caption{Grouping of image pixels in level 3 to introduce 2D recursion.}
	\label{fig-Illustration_Grouping}
\end{figure}

\begin{figure*}	
	\centerline{\includegraphics[width=1\textwidth]{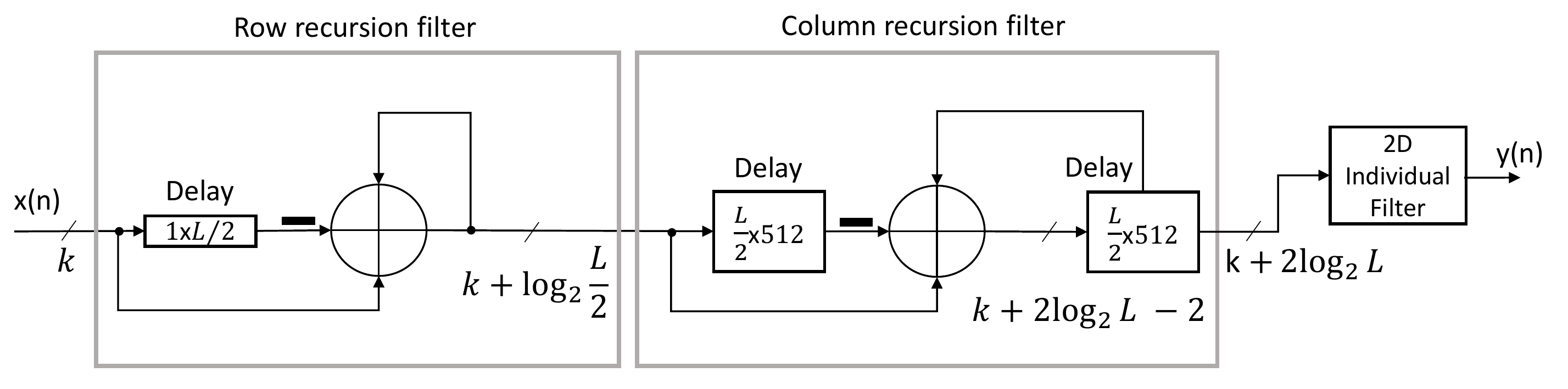}}
	\caption{Illustration of 2D recursive UWT for a single wavelet channel.}
	\label{fig-2D_RUWT}
\end{figure*}

Now let's examine the 2D filters. To illustrate recursion, we consider the kernel for HH sub-band in level 3 which is given in Fig.~\ref{fig-2D_Kernel_HH_Lev3}. 
Exploiting the structure of the filter, we can group the image pixels as shown in Fig.~\ref{fig-Pixel_Grouping_Lev3}, where A, B, C and D represent the sum of pixels in the group. Now, HH sub-band can be calculated as 
\begin{equation}
\text{HH}= \text{D}+\text{A}-\text{C}-\text{B},
\label{eq_2D_Recursion_HH}			
\end{equation}
which is same as \eqref{eq-2D}. Here, the values A, B, C and D can be found by applying a low-pass filter of size $\frac{L}{2}$x$\frac{L}{2}$ on the input. This low-pass filter can be implemented by two 1D recursions given by \eqref{eq_1D_recursion_L_Filter} with length of the filter as $\Lbtwo$ which results in four additions. Once the low pass image is found, all the sub-bands can be calculated using \eqref{eq-2D}.  Hence, the total number of additions required to find a sub-band is $4+3 = 7$. The first four additions are common for all the sub-bands, hence the average additions per output pixel is (4+3$\times$4)/4 = 4. Implementation of the filter for a single wavelet channel is shown in Fig.~\ref{fig-2D_RUWT}. Here we need one $\Lbtwo$ and two $\LbtwoX$ buffers which is less than the 1D recursive implementation. The bit-width requirements for these buffers are also less.
As already discussed in the previous section, the number of additions will increase in the recombination side. For 1D implementation the average additions will be 5 and for 2D, it will be 7.

We call the method of recursive implementation as recursive UWT (RUWT) for further reference. Table \ref{tab-Average_Additions} summarizes the average additions for filtering level five. Since the delays are of large size, they are implemented using Block RAMs (BRAM) available in FPGA. Table \ref{tab-Usage_36K_BRAM} gives the memory usage in terms of number of 36K BRAMs in the fifth level of filtering. We have chosen level-five for comparison since it has the maximum filter length across all the levels, which is equal to 32.
\begin{table}[h]
	\centering	
	\caption{Average additions per output pixel for filtering level five in different methods of implementing UWT}
	\label{tab-Average_Additions}
	\begin{tabular}{l S[table-format=3.2] S[table-format=3.2]}
		\hline\\
		\textbf{Method}   & \textbf{Decomposition}      & \textbf{Recomposition}     	\\ \hline\\ 
		{\color[HTML]{000000} 1D UWT}                           & 1.50                        & 6.00                     	\\ 
		2D UWT                                                  & 2.00                        & 12.00                    	\\ 
		1D Convolution\tablefootnote{Using combined filters} \label{foot_Conv}	
		& 46.50                       & 62.00                    	\\
		2D Convolution\footnotemark[1]            				& 1023.00                     & 1023.00                 	\\ 
		1D RUWT                                                 & 3.75                        & 5.00                     	\\ 
		2D RUWT                                                 & 4.00                        & 7.00                     	\\ \hline
	\end{tabular}	
\end{table}
\begin{table}[h]
	\centering
	\caption{Number of Block RAMs (36K) required for filtering level five in different methods of implementing UWT. Here we assume a bit-width of 16 for calculating the utilization.}
	\label{tab-Usage_36K_BRAM}
	\begin{tabular}{l c c}
		\hline\\		
		\textbf{Method}                        		& \textbf{Decomposition}      & \textbf{Recomposition}      \\ \hline\\
		1D UWT                                      & 7.5                & 15                 \\ 
		2D UWT                                      & 4                  & 15                 \\ 
		1D RUWT                                     & 17                 & 32                 \\ 
		2D RUWT                                     & 15                 & 30                 \\ \hline
	\end{tabular}
\end{table}

We now select any one of the discussed methods or a combination of them for the final implementation. UWT has the lowest number of additions among all. However, we combine all the resulting sub-bands from a level to a single low pass band during recombination. But to find the shrinkage factors, we need all of them separately as shown in Fig.~\ref{fig-Gen_RE_2_Level}. Hence, UWT methods can not be used in recombination side. This narrows down the list to UWT or RUWT for decomposition and RUWT for recomposition. UWT has lesser number of computations compared to RUWT. So we chose it for decomposition. We also have to make a choice between 1D and 2D implementations. Since 2D methods have lesser memory requirements we chose them. Finally, we select 2D UWT for decomposition and 2D RUWT for recomposition.
\subsection{Sub-band time alignment}
We find the sixteen wavelet sub-bands which constitute the columns of the matrix $\Cpsi$ in \eqref{eq-alpha} in parallel. The results will not be time aligned since the kernel sizes are different.
\begin{figure}[h!]	
	\centerline{\includegraphics[width=0.8\columnwidth]{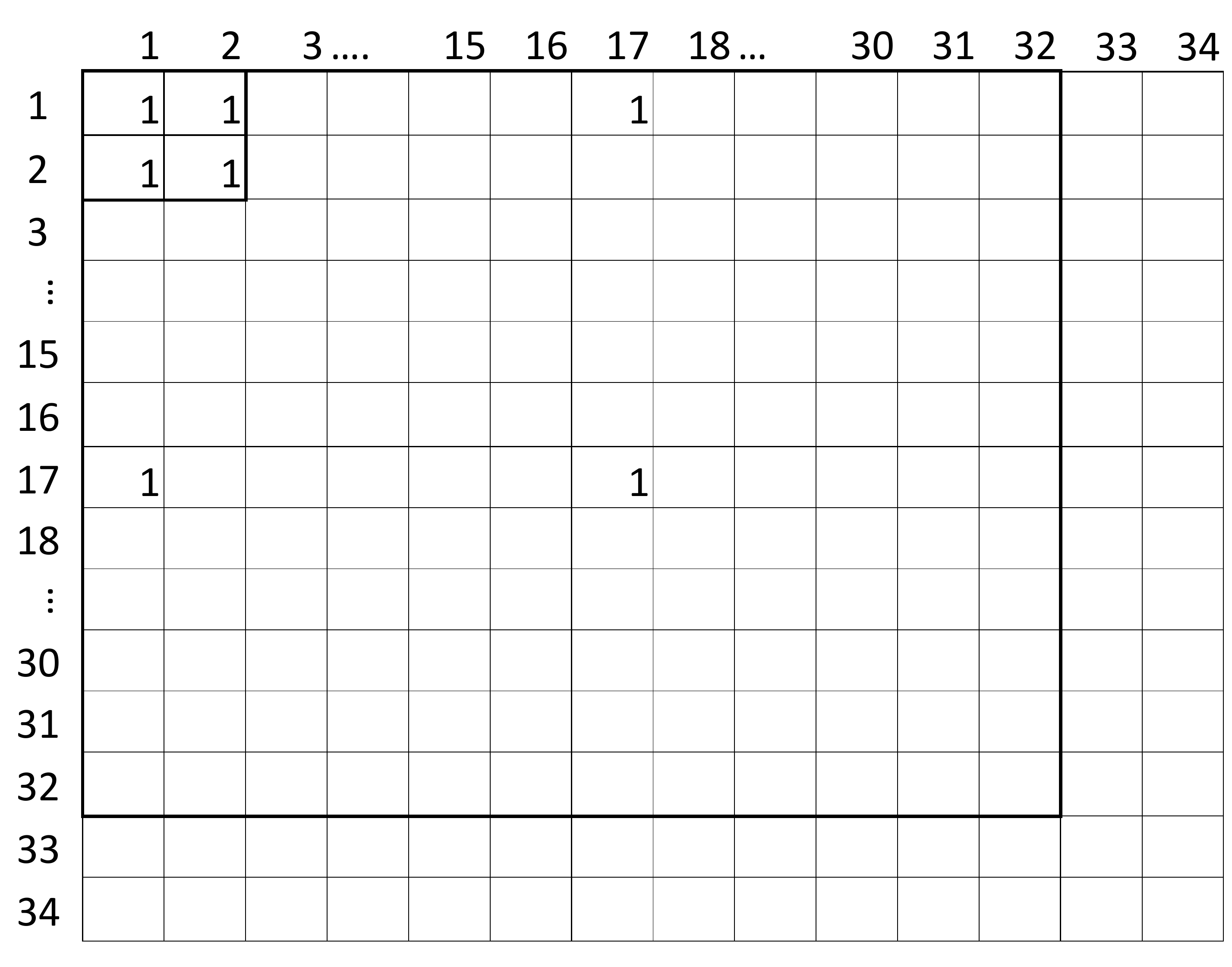}}
	\caption{When $\GG$ kernels in level 1 and 5 operate in parallel on an image.}
	\label{fig-Time_Misalign}
\end{figure}
For example, if we consider the parallel application of filters in first and fifth level on an image as shown in Fig.~\ref{fig-Time_Misalign}, the output of the decomposition kernel in level one is available along with the second row, third element of the input image, whereas the output from a filter in the fifth level will only be available with the $16^\text{th}$ row,  $17^\text{th}$ element. Thus there is a delay of $16\times512 + 16$ clock cycles between the corresponding elements of sub-bands in the first and last level. Similar delays exist in other levels too. However, to estimate the shrinkage factors in a pipelined fashion, we need the sub-bands to be time aligned. This can be achieved by delaying the inputs to the filters in levels one to four so that outputs from all the levels are time aligned. This can also be done by delaying the outputs which is not preferred since the output bit-width is more which leads to increased memory utilization. However, there is a practical difficulty in implementing this on the decomposition side as the input to a particular level is the low pass output from the previous level. 
\begin{figure}[h!]	
	\centerline{\includegraphics[width=0.75\columnwidth]{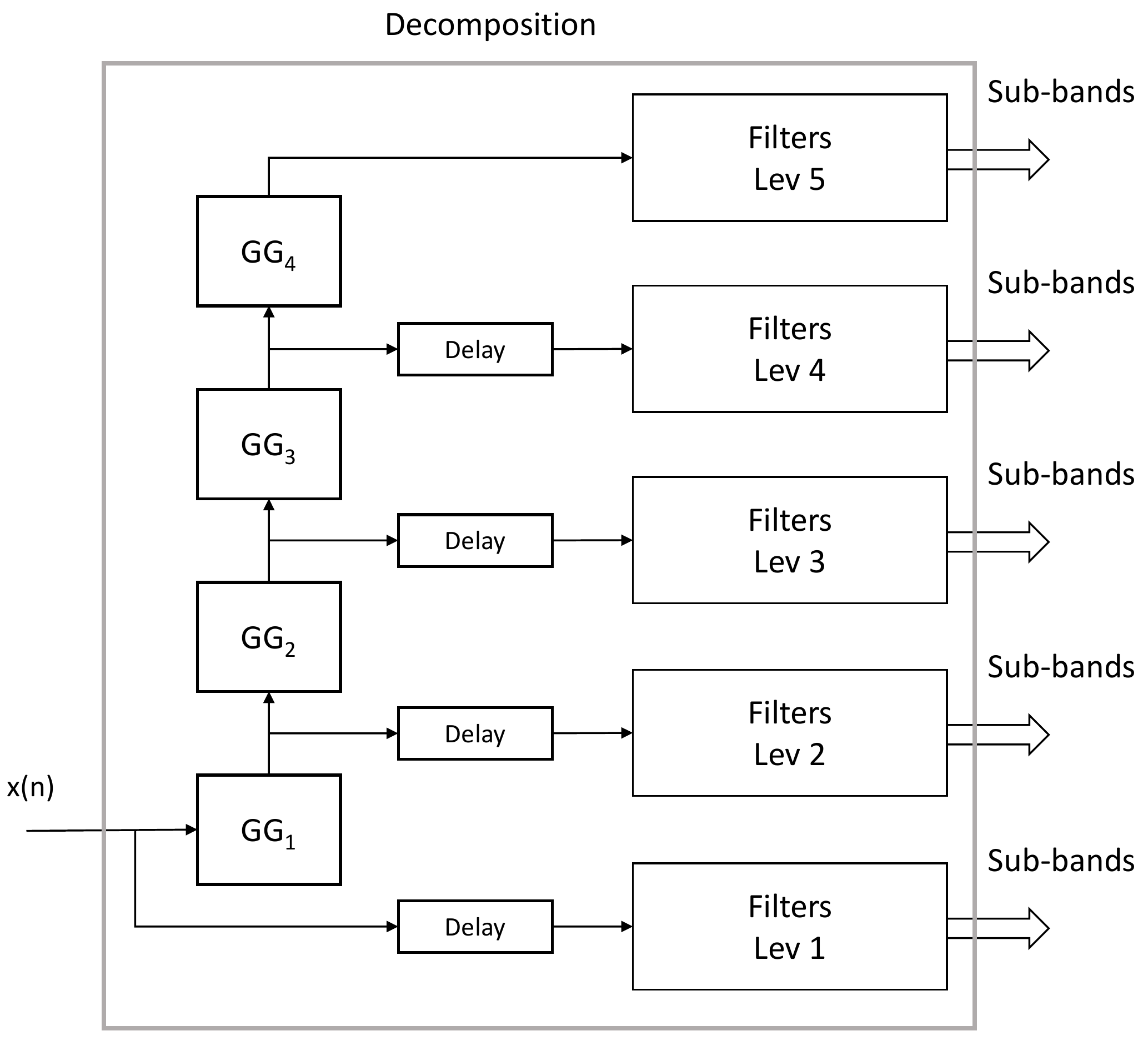}}
	\caption{Advance calculation of LL sub-bands in order to achieve time alignment.}
	\label{fig-Time_align_Decomp}
\end{figure}
Now, if we delay the input to a particular level, inputs to the higher levels will also get delayed by the same amount. Fig.~\ref{fig-Time_align_Decomp} illustrates one way of addressing this. Here we find the low pass sub-bands of levels one to four in advance. These intermediate results are then properly delayed and applied to the decomposition filters so that their outputs are time aligned.
\subsection{Bit-width analysis and number format}
We prefer fixed point implementation as it’s size, power consumption and memory usage is less compared to floating point and can achieve more speed of operation~\cite{Xilinx}. The input pixels of the image are eight bit unsigned numbers and the corresponding Q format is Q8.0. Since the filtering involves equal number of additions and subtractions, the range of the result from decomposition filters after scaling will be $[-\frac{M}{2}, \frac{M}{2}]$, where $M = \max|I|$ and $I$ is the input pixel value. Hence the integer part of the result can be represented in 7 bits and the extra bit becomes the sign bit. The LL sub-band on the other hand have only additions, hence the output after scaling will be unsigned with the same range as input and we need eight bits to represent the integer part. The number of bits in the fractional part is depended on the filter size and thus on the level of filtering.  In the case of recomposition filters, even though we have equal number of additions and subtractions, the input to the filters are signed numbers. Here the integer part of the input is represented by seven bits and hence same number of bits are required to represent the integer part of the output as well. Table \ref{tab-Q_Format_Original} summarizes the change in Q format over different filtering levels except for LL sub-band. For LL sub-bands the Q format is similar except that the integer part is represented by 8 bits and also is unsigned.
\begin{table}[h]
	\centering
	\caption{Change of Q format for different sub-band outputs except for LL. Input format is Q8.0(unsigned) and the outputs are signed.}
	\label{tab-Q_Format_Original}
	\scalebox{0.9}{%
		\begin{tabular}{c c S[table-format=3.2] S[table-format=3.2]} 
			\hline
			\textbf{Level}&\textbf{Scaling($\frac{1}{L})$}&\textbf{Qm.n(Decomposition)}&\textbf{Qm.n(Recomposition)}\\
			\hline\\
			1 &$2^{-2}$  	&Q7.2	&Q7.4\\
			2 &$2^{-4}$  	&Q7.4	&Q7.8\\
			3 &$2^{-6}$ 	&Q7.6	&Q7.12\\	
			4 &$2^{-8}$  	&Q7.8	&Q7.16\\
			5 &$~2^{-10}$	&Q7.10	&Q7.20\\	
			\hline	
	\end{tabular}}
\end{table}

The bit-width at the output of the recombination filter varies from 12 bits at the first level to 28 bits at the final level. We don't need to keep all these bits because during shrinkage we scale each of these sub-bands with $\alpha_i$ which is always near unity or less and then combine and round to the nearest eight bit integer number. Thus the fractional parts at farthest positions will have less effect on the result. Hence we truncate the fractional positions to eight bits as given in Table \ref{tab-Q_Format_reduced} so that we have 16 bit numbers at the maximum.\\

\begin{table}[h]
	\centering
	\caption{Q format after truncation.}
	\label{tab-Q_Format_reduced}
	\scalebox{1.0}{%
		\begin{tabular}{c S[table-format=3.2] S[table-format=3.2]} 
			\hline
			\textbf{Level}&\textbf{Qm.n(Decomposition)}&\textbf{Qm.n(Recomposition)}\\
			\hline\\
			1  	&Q7.2	&Q7.4\\
			2 	&Q7.4	&Q7.6\\
			3 	&Q7.6	&Q7.6\\	
			4	&Q7.6	&Q7.6\\
			5 	&Q7.6	&Q7.6\\	
			\hline	
	\end{tabular}}
\end{table}
\section{Estimation of Shrinkage Parameter}
There are four matrix operations involved in the estimation of shrinkage parameter vector $\alpha$ given in \eqref{eq-alpha}; a matrix multiplication, inversion of resulting matrix, a matrix vector multiplication and a vector addition. Since we use a five level wavelet decomposition, we will have sixteen wavelet sub-bands which constitute the columns of the matrix $\Cpsi$. We calculate all these sixteen sub-bands in parallel and thus a row of this matrix becomes available in each clock. We now write the matrix multiplication in \eqref{eq-alpha} as an outer product
\begin{equation}
\mathbf{Q} = (\Cpsi^\text{T}\Cpsi) = \Cphi\Cphi^\text{T}	
\label{alpha}  
\end{equation}
where $\Cphi = \Cpsi^\text{T}$. Now the consecutive columns of the matrix $\Cphi$ become available in successive clocks. The outer product of a matrix can be found as summation of outer products of its columns~\cite{meyer2000matrix}. This outer product generation and accumulation can be realized using the DSP cores available in FPGA. Since each of these columns have sixteen elements, there are 256 parallel multiplications in the generation of outer products. In this, 120 multiplications are redundant since $\mathbf{Q}$ is symmetric. So we find the upper diagonal entries of $\mathbf{Q}$ first and then reflect it along the diagonal to find the complete matrix. We need only 130 DSP cores to do this. We need another 16 DSP cores to do the matrix vector multiplication in \eqref{eq-alpha} and the vector addition can be implemented using LUTs. We now rewrite \eqref{eq-alpha} as
\begin{equation}
\alp^* = \mathbf{Q}^{-1}\mathbf{c}	
\label{eq-alpha_GD}  
\end{equation}
where c = $(\Cpsi^\text{T}\yy - \qq)$. The vector $\alp^*$ is the solution of the linear equation
\begin{equation}
\mathbf{Q}\alp^*= \mathbf{c}	
\label{eq-Linear_Eq}  
\end{equation}
Since $\QQ$ is defined as $(\Cpsi^\text{T}\Cpsi)$, it is a positive semidefinite matrix. But in our denoising framework it can be considered as a positive definite matrix as we assume the signal is corrupted by noise which is uncorrelated. Hence we can use convex optimization techniques to solve this equation instead of actually inverting the matrix to find the solution, which require more resources. We use gradient descent algorithm to solve \eqref{eq-Linear_Eq} since it has lesser number of computations compared to other optimization techniques~\cite{bubeck2015convex}. Here the weight update equation can be written as
\begin{equation}
\alp_i = \alp_{i- 1} + \mu(\mathbf{c} - \QQ\alp_{i- 1})
\label{GD}
\end{equation}
The matrix vector multiplication in \eqref{GD} can be implemented using 16 DSP cores. Initial weight vector $\alp_0$ is taken as one, since it is the ideal solution if there is no noise. The step size is chosen as $\frac{1}{2^{13}}$ which is implemented using shift operation.

\section{Hardware Implementation}
\begin{figure}[h!]
	\centerline{\includegraphics[width=1\columnwidth]{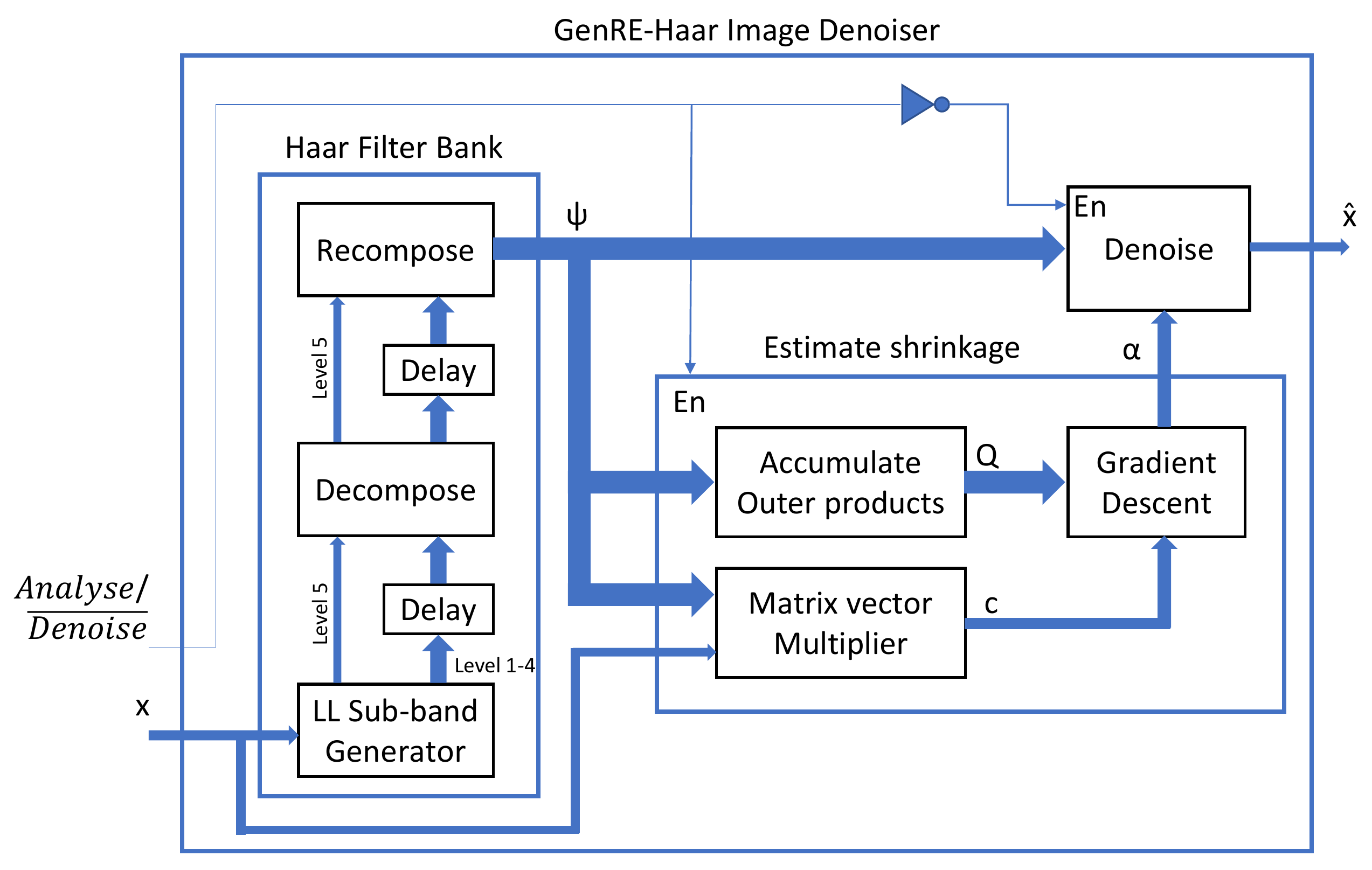}}
	\caption{Hardware architecture of GenRE-Haar Image denoiser. $\xx$ and $\hat{\xx}$ represents row vectorized input and output images respectively.}
	\label{fig-GenRE_Arch}
\end{figure}

The different blocks described in the previous section are now combined to give the final architecture of the GenRE-Haar Image denoiser which is shown in Fig.~\ref{fig-GenRE_Arch}. Since we assume $512\times512$ images, the matrix $\Cpsi$ will have a size $262144\times16$. This matrix is first used to find the shrinkage $\alp$ and then to denoise the image. Hence we have to store this until the shrinkage factor is calculated. Block RAMs can not be used for this purpose since the matrix size is too large. To address this we apply the input image twice. In the first iteration we calculate and store the shrinkage factors. During the next iteration we denoise the image using these stored shrinkage factors. The $\mathbf{Anlyse}/\overline{\mathbf{Denoise}}$ input serves this purpose. The performance of an FPGA design is determined by its resource utilization and critical path delay which determines the frequency of operation. Table \ref{tab-Utilization} gives the resource utilization and maximum operating frequency of the entire design. It has been observed that the improvement in signal to noise ratio for full precision implementation is negligible compared to 16 bit implementation.
\begin{table}[]
	\centering
	\caption{Post implementation resource utilization of GenRE-Haar denoiser. The precision mentioned here refers to the precision of Haar filter bank.}
	\label{tab-Utilization}
	\begin{tabular}{lll}
		\hline
		Resources & Full precision & 16 bit precision \\ \hline\\
		\#LUT     & 30516          & 18397            \\
		\#FF      & 41349          & 23549            \\
		\#BRAM36  & 150            & 132              \\
		\#DSP     & 168            & 168              \\ 
		FMax(MHz) & 169            & 183              \\ \hline
	\end{tabular}
\end{table}
\section{Results}
\begin{figure}
	\begin{subfigure}[t]{0.25\columnwidth}\centering
		\includegraphics[width=0.95\columnwidth]{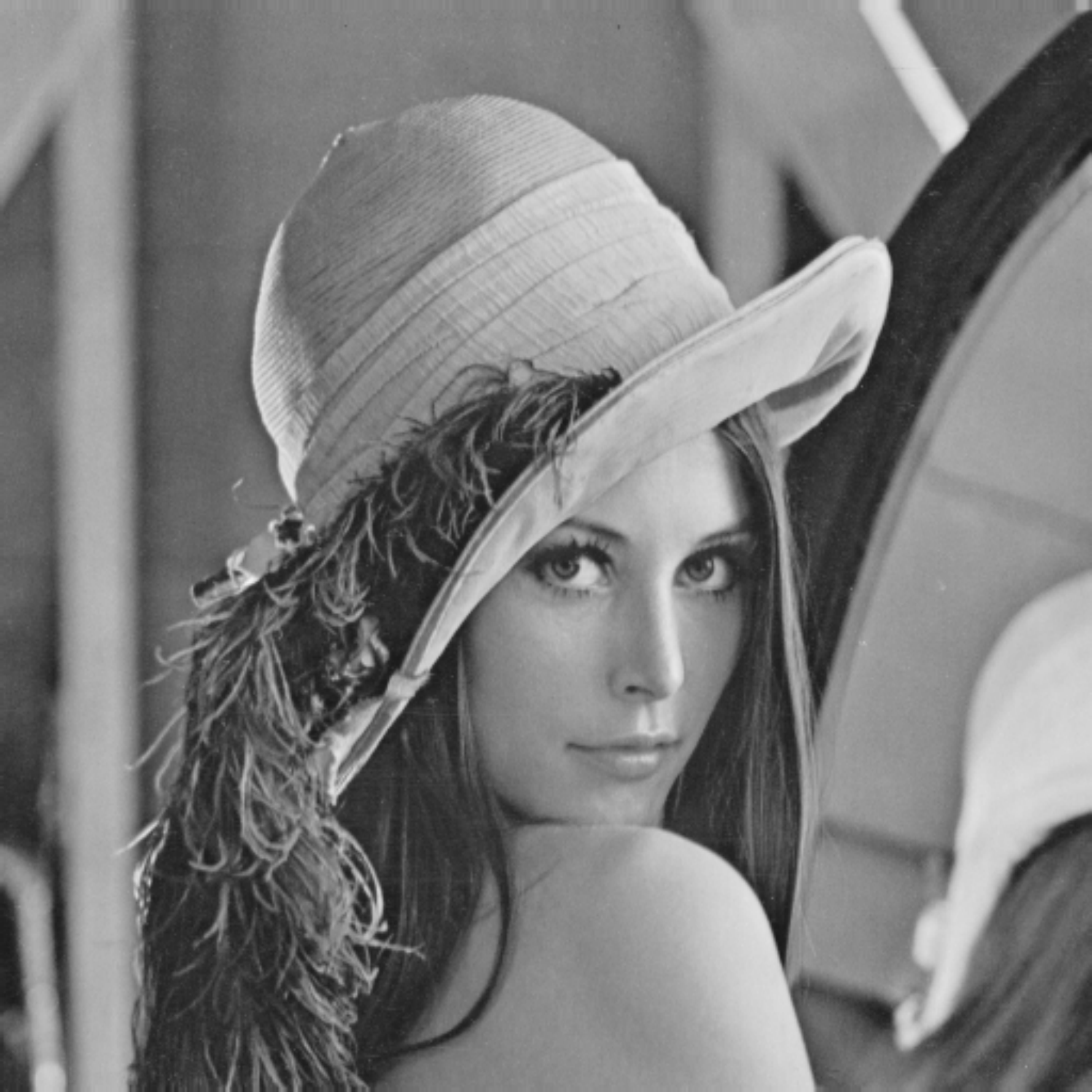}
		\caption{Clean Image.}
	\end{subfigure}%
	\begin{subfigure}[t]{0.25\columnwidth}\centering
		\includegraphics[width=0.95\columnwidth]{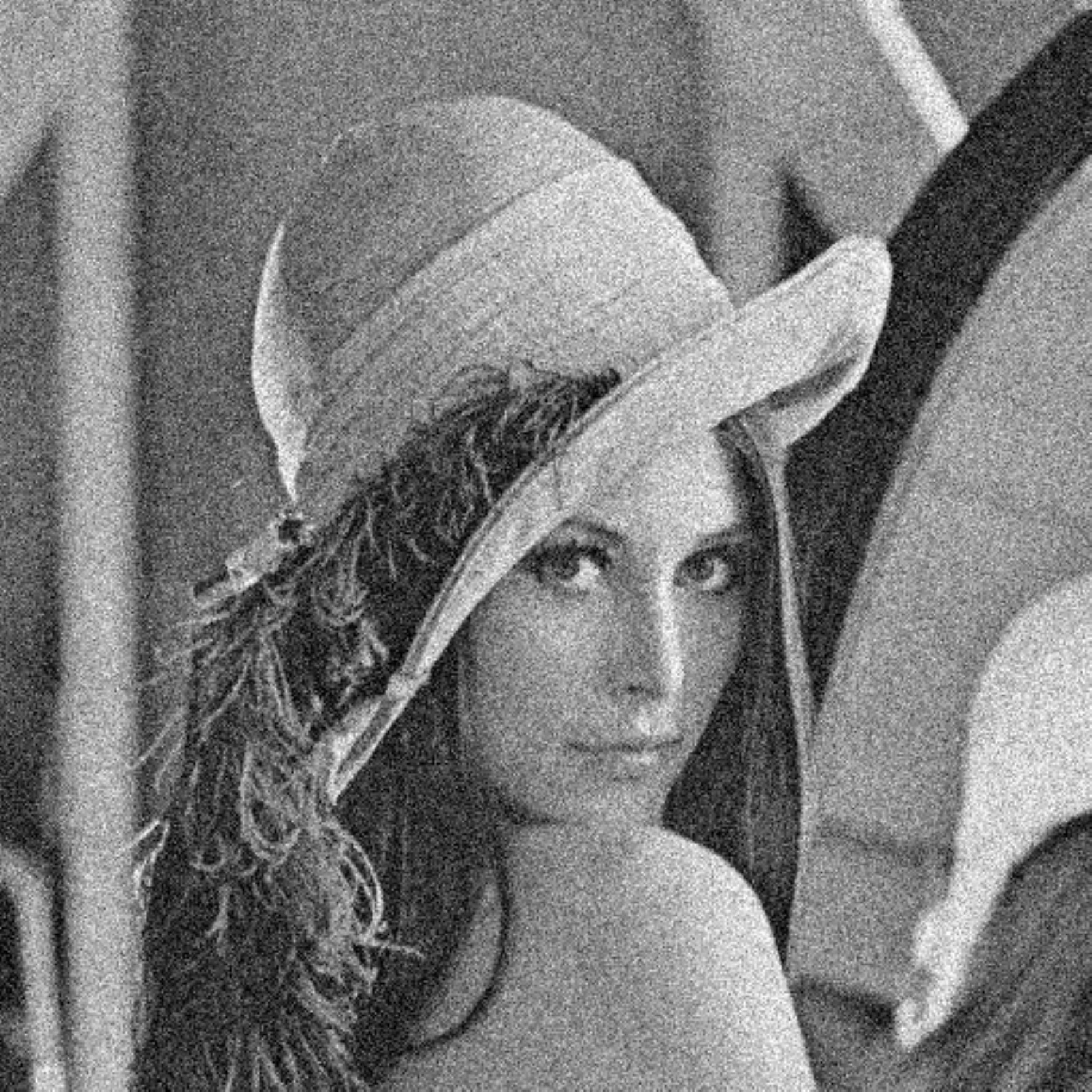}
		\caption{Noisy Image.\\PSNR =\\20.246 dB.}
	\end{subfigure}%
	\begin{subfigure}[t]{0.25\columnwidth}\centering
		\includegraphics[width=0.95\columnwidth]{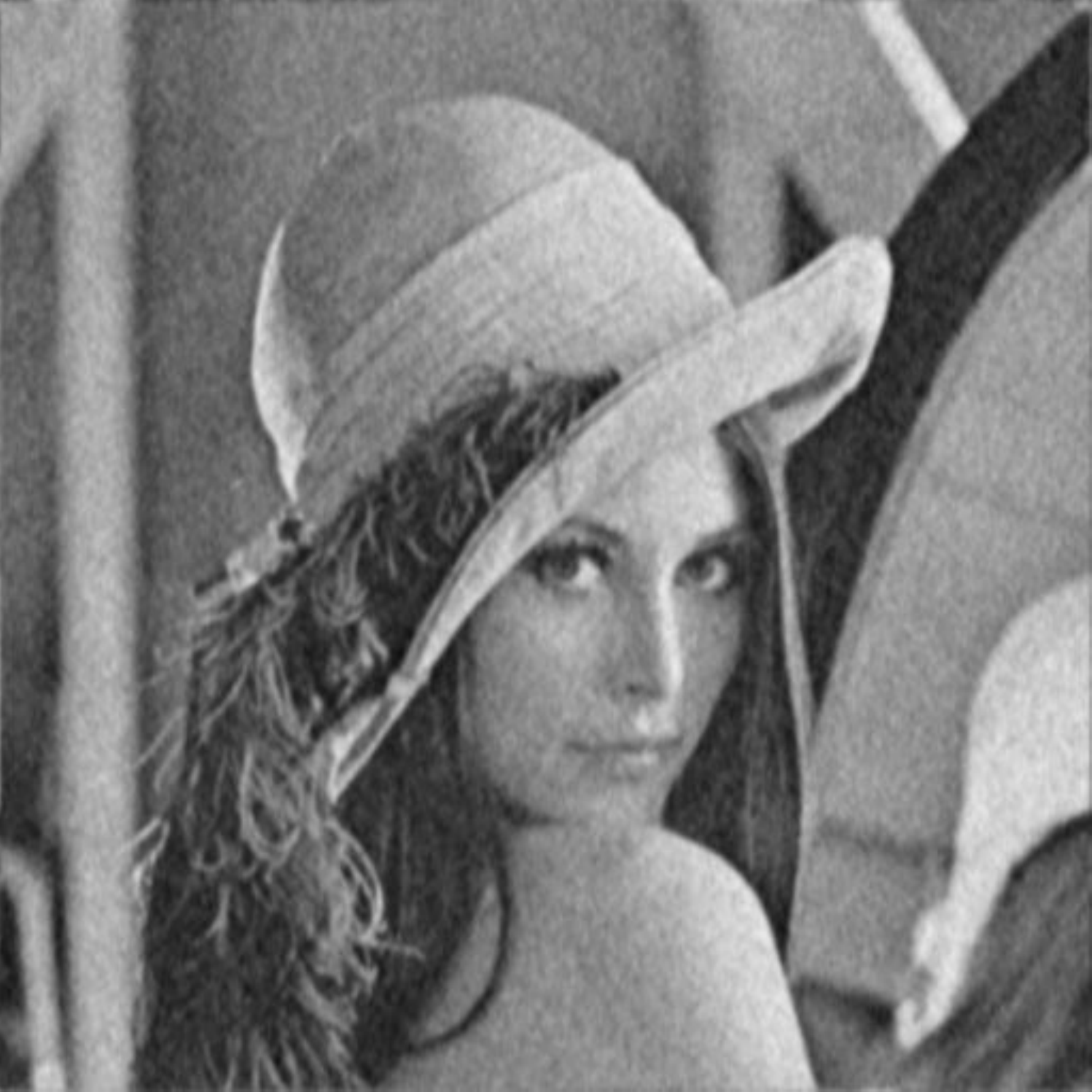}
		\caption{MATLAB\\reference\\PSNR =\\29.204 dB.\\PSNR Gain =\\8.958.}
	\end{subfigure}%
	\begin{subfigure}[t]{0.25\columnwidth}\centering
		\includegraphics[width=0.95\columnwidth]{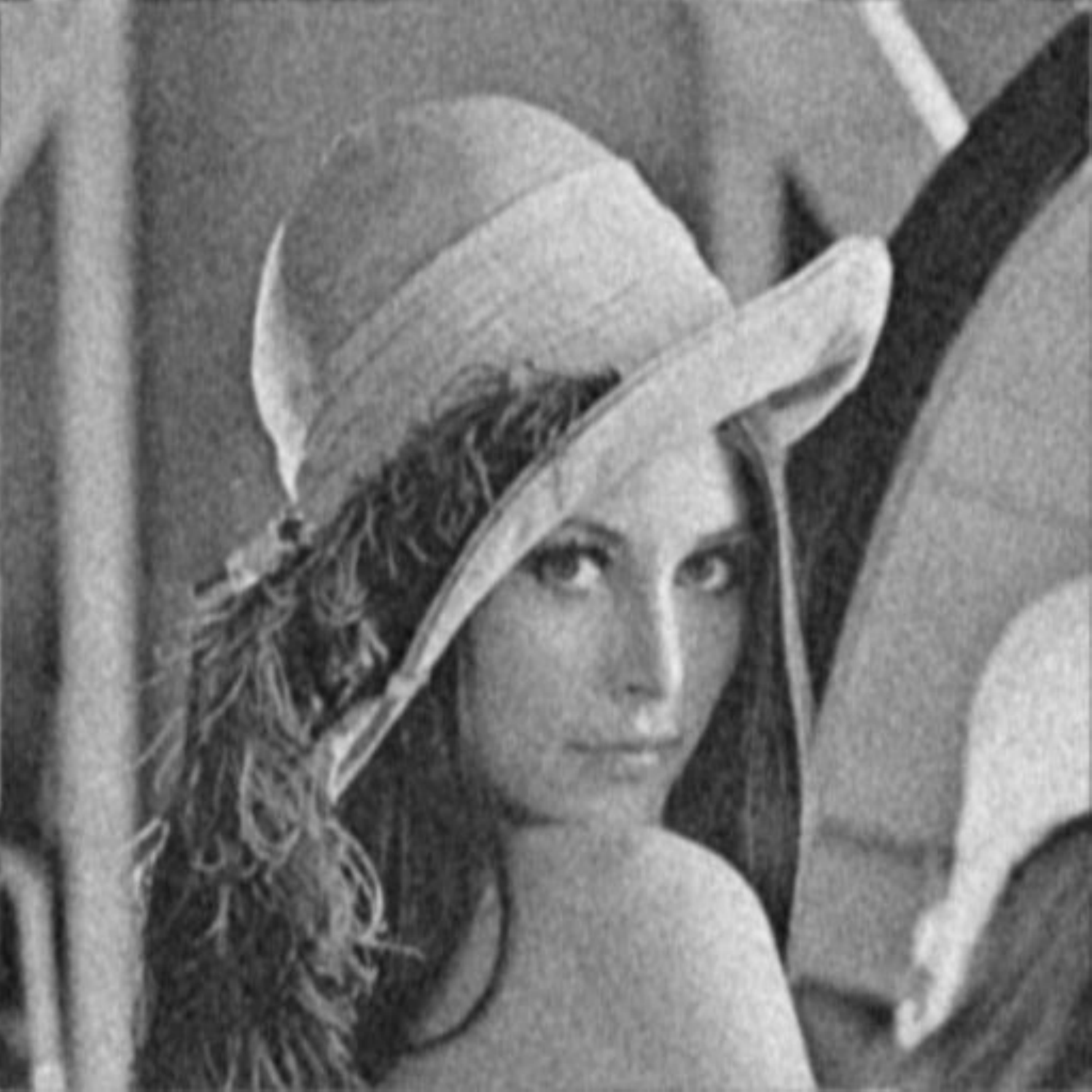}
		\caption{FPGA result.\\PSNR =\\29.110 dB. \\PSNR Gain =\\8.864 dB.}
	\end{subfigure}	
	\begin{subfigure}[t]{0.25\columnwidth}\centering
		\includegraphics[width=0.95\columnwidth]{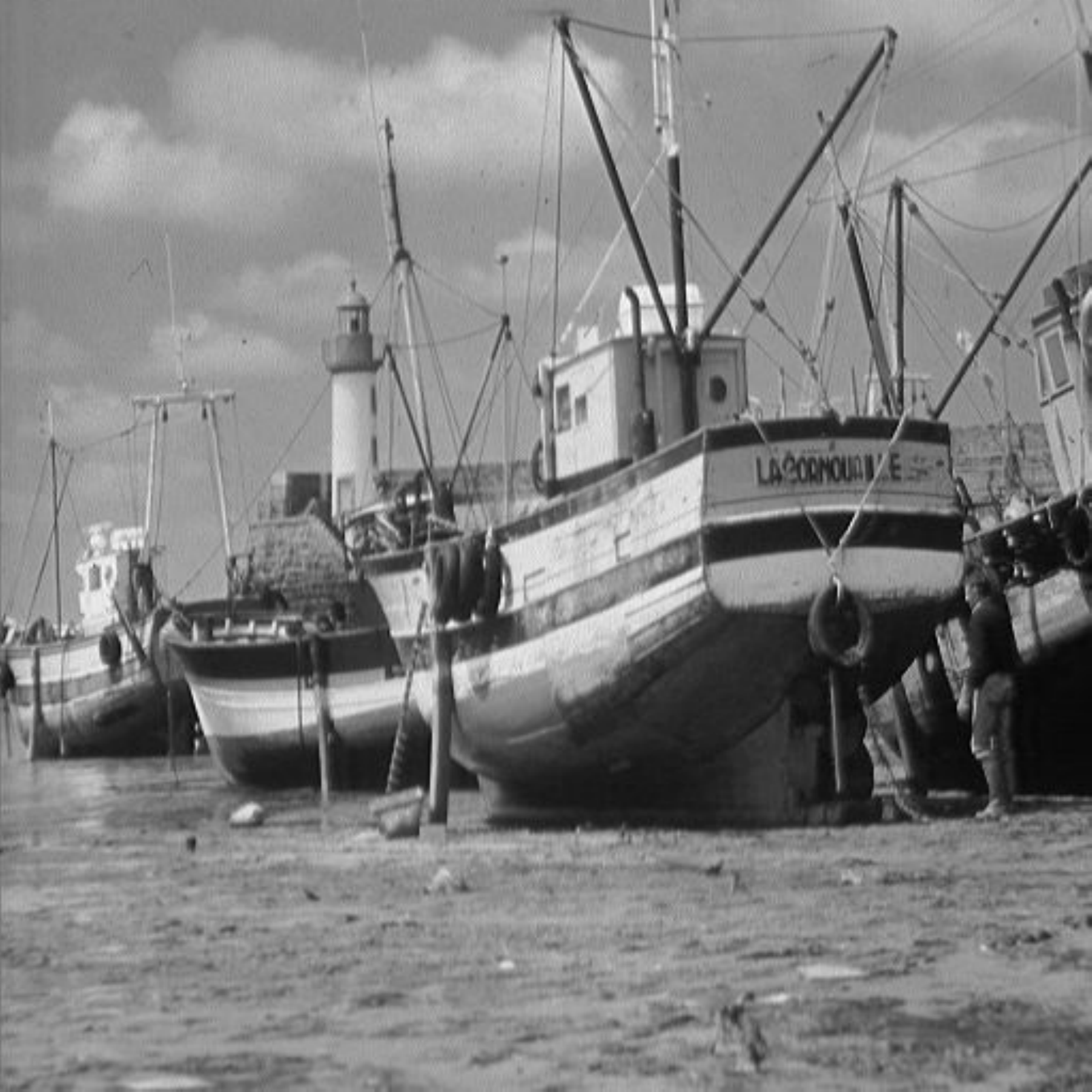}
		\caption{Clean Image.}
	\end{subfigure}%
	\begin{subfigure}[t]{0.25\columnwidth}\centering
		\includegraphics[width=0.95\columnwidth]{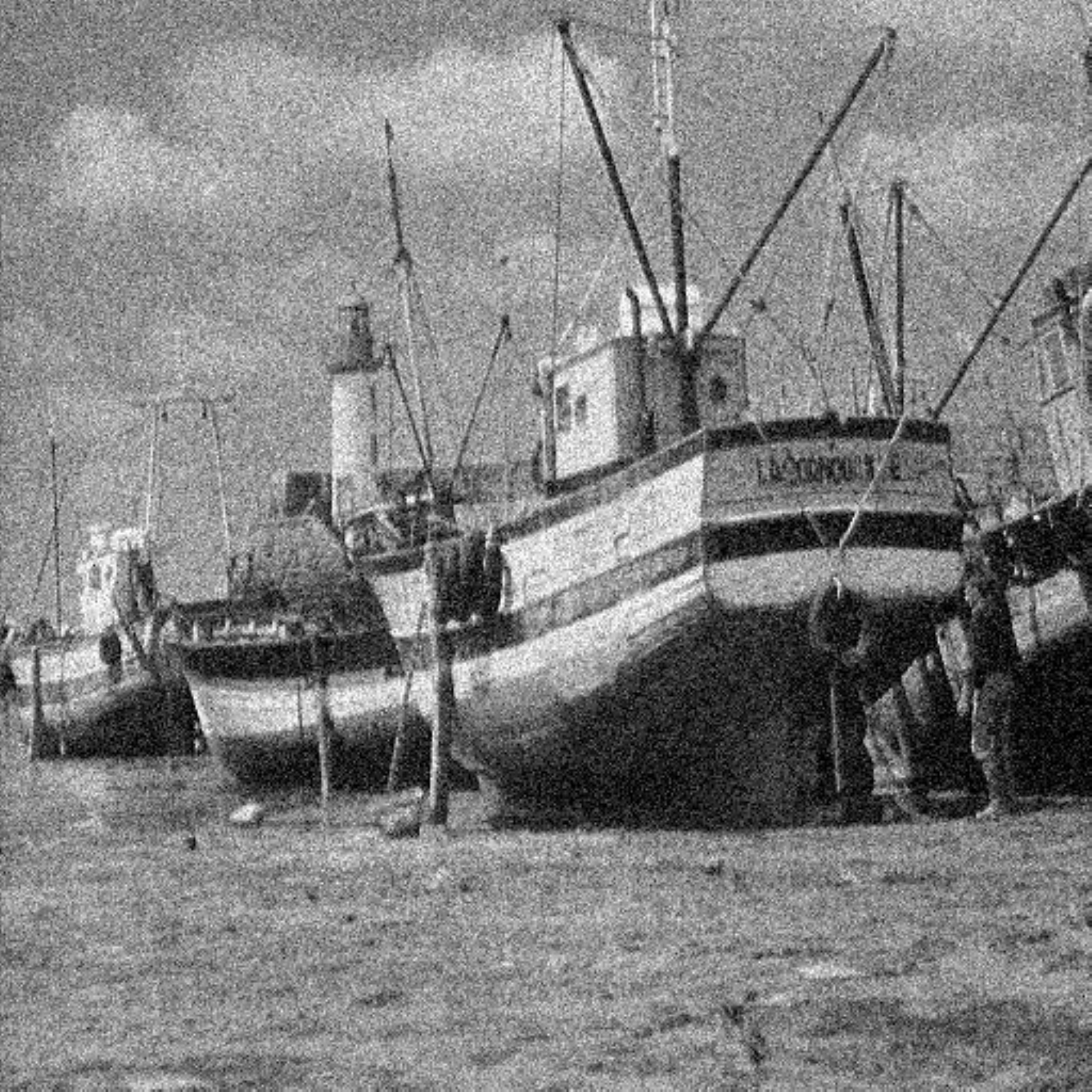}
		\caption{Noisy Image.\\ PSNR =\\20.268 dB.}
	\end{subfigure}%
	\begin{subfigure}[t]{0.25\columnwidth}\centering
		\includegraphics[width=0.95\columnwidth]{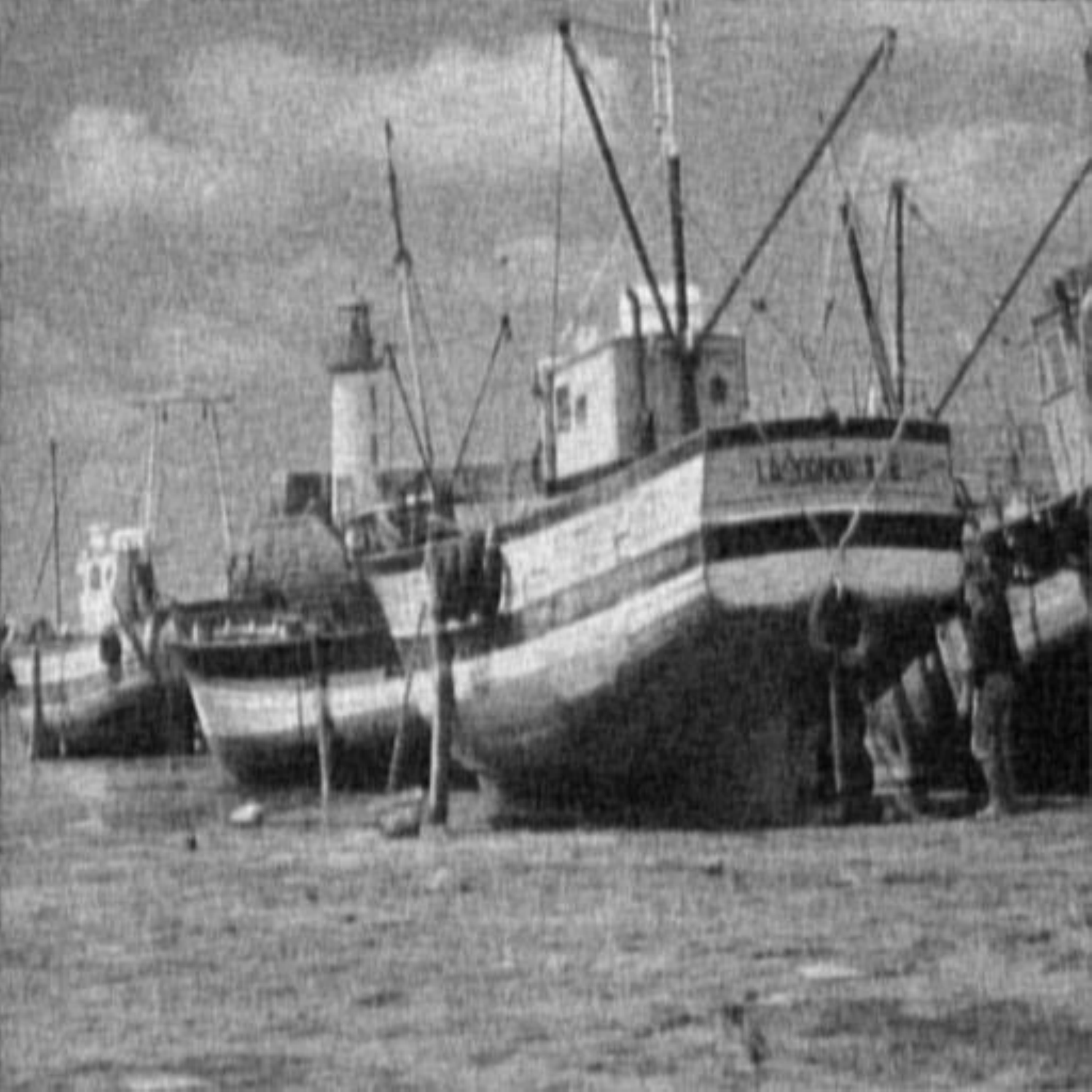}
		\caption{MATLAB\\reference.\\ PSNR =\\27.621 dB.\\PSNR Gain =\\7.353 dB.}
	\end{subfigure}%
	\begin{subfigure}[t]{0.25\columnwidth}\centering
		\includegraphics[width=0.95\columnwidth]{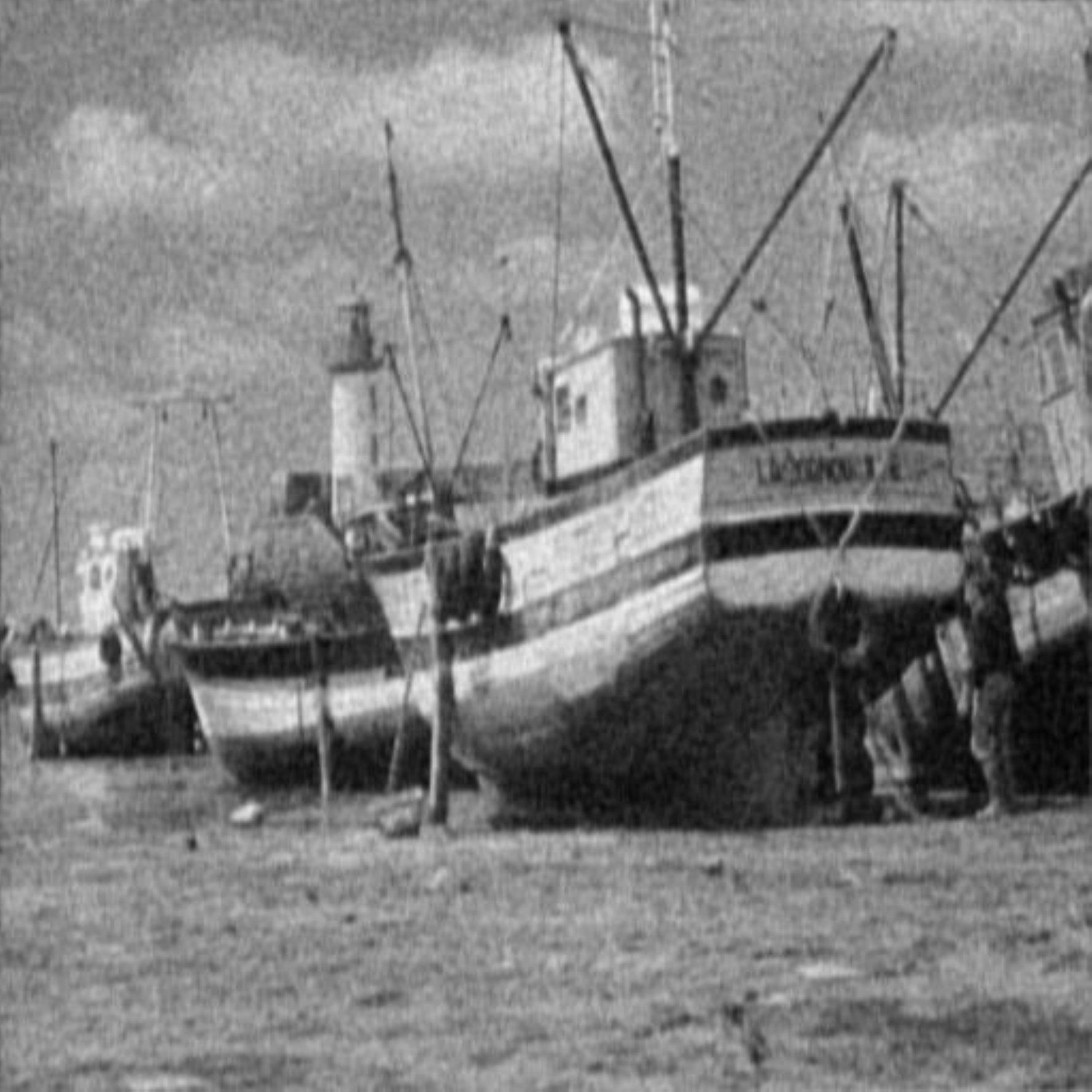}
		\caption{FPGA result.\\PSNR =\\27.521 dB.\\PSNR Gain =\\7.253 dB}
	\end{subfigure}
	\begin{subfigure}[t]{0.25\columnwidth}\centering
		\includegraphics[width=0.95\columnwidth]{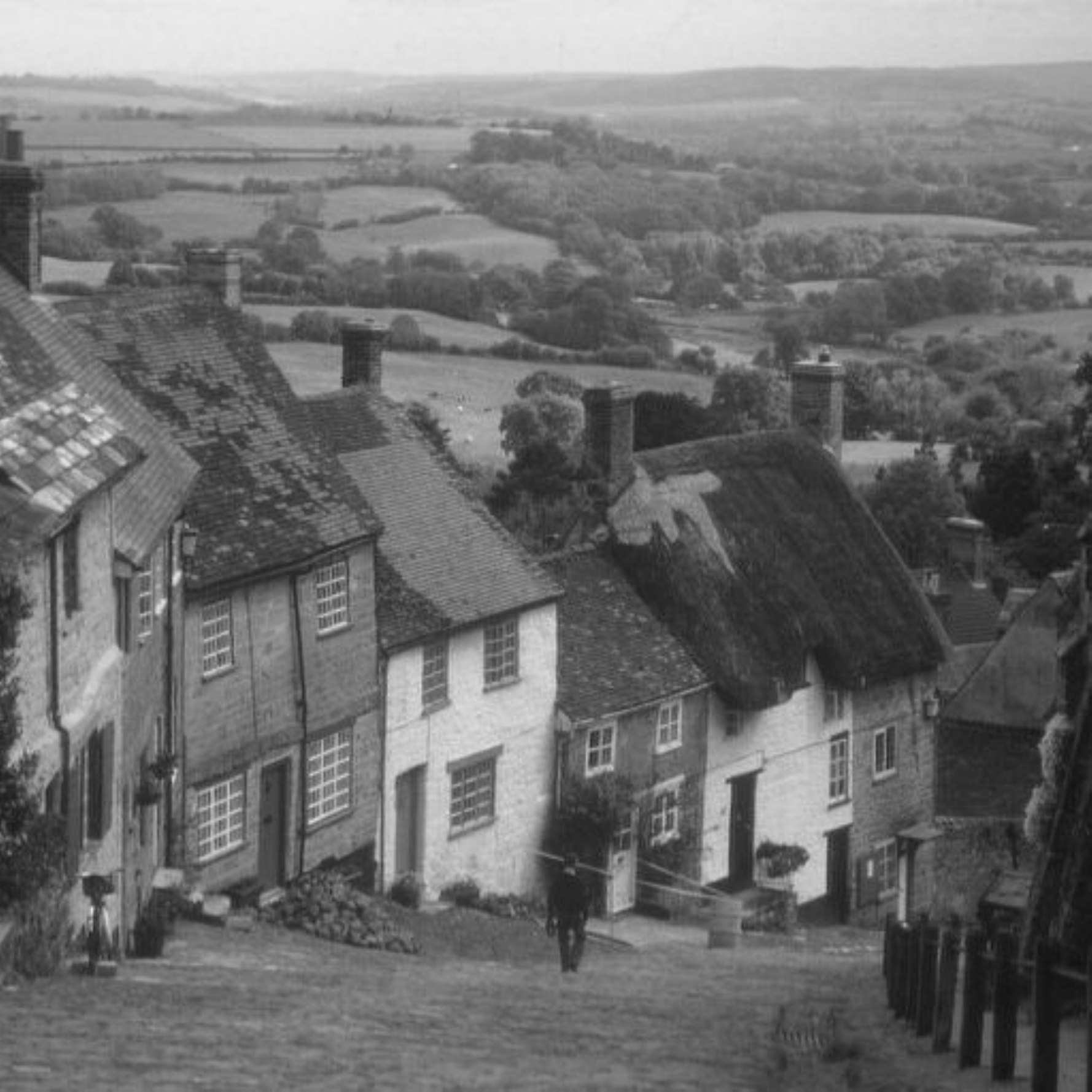}
		\caption{Clean Image.}
	\end{subfigure}%
	\begin{subfigure}[t]{0.25\columnwidth}\centering
		\includegraphics[width=0.95\columnwidth]{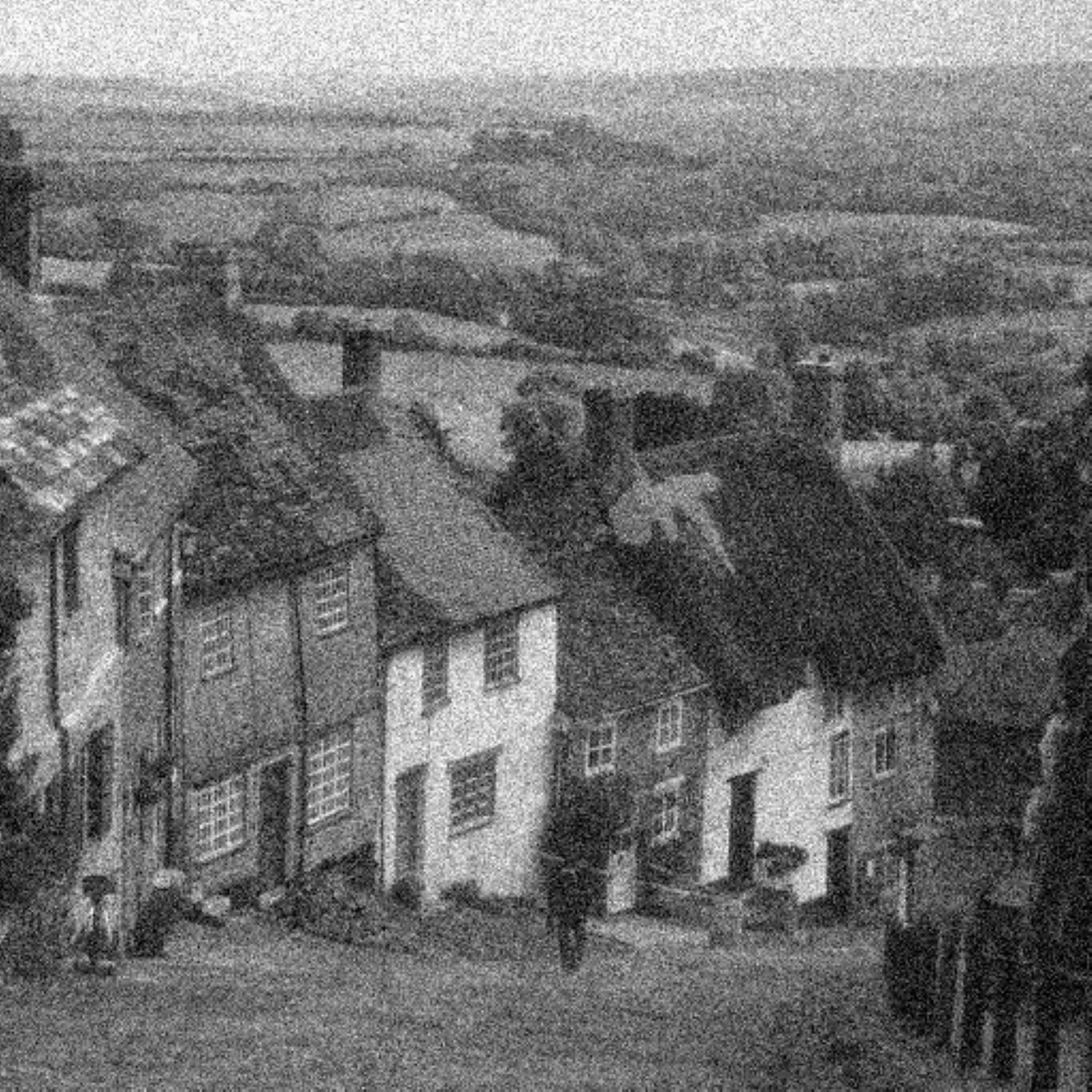}
		\caption{Noisy Image.\\PSNR =\\20.236 dB.}
	\end{subfigure}%
	\begin{subfigure}[t]{0.25\columnwidth}\centering
		\includegraphics[width=0.95\columnwidth]{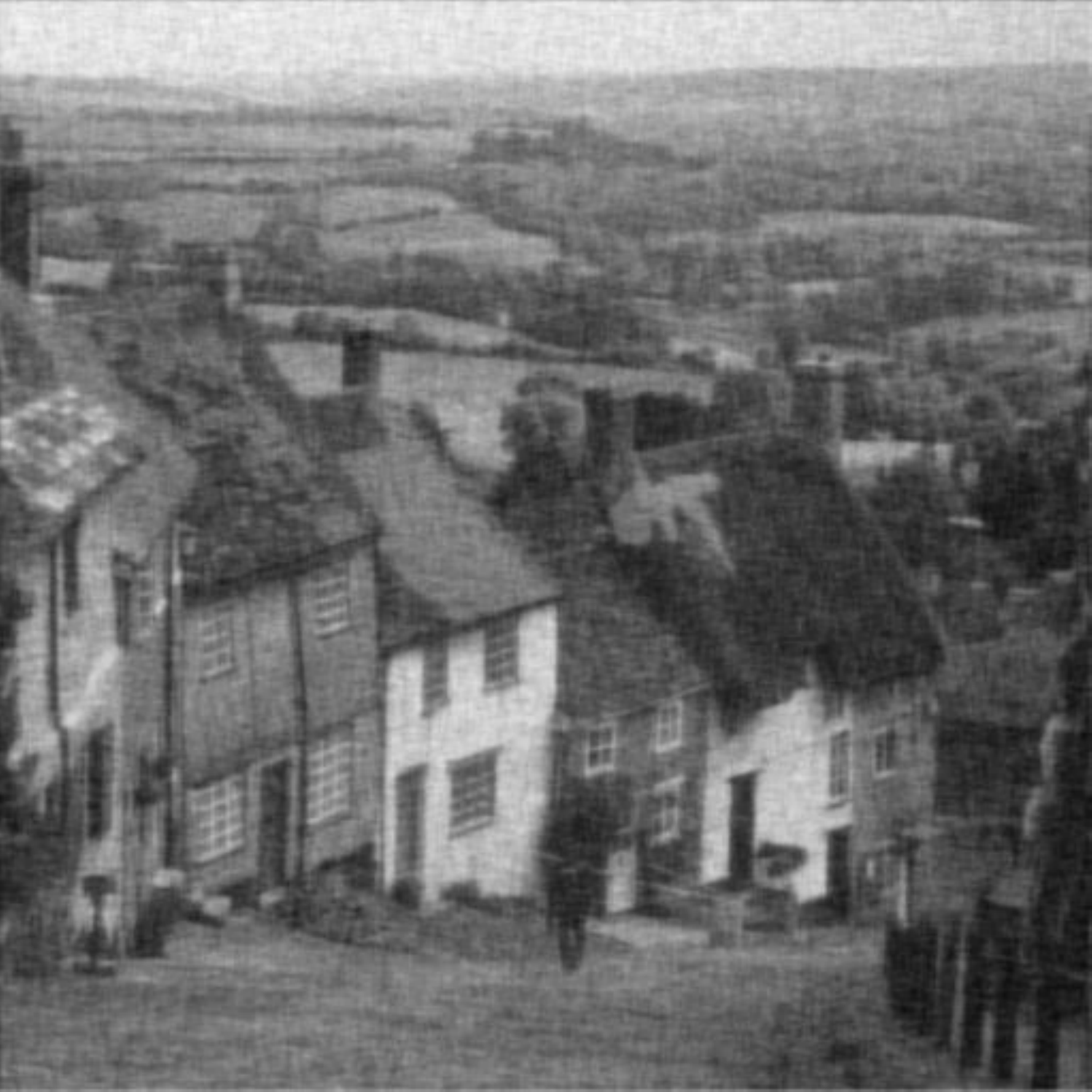}
		\caption{MATLAB\\reference.\\PSNR =\\28.194 dB.\\PSNR Gain =\\7.958.}
	\end{subfigure}%
	\begin{subfigure}[t]{0.25\columnwidth}\centering
		\includegraphics[width=0.95\columnwidth]{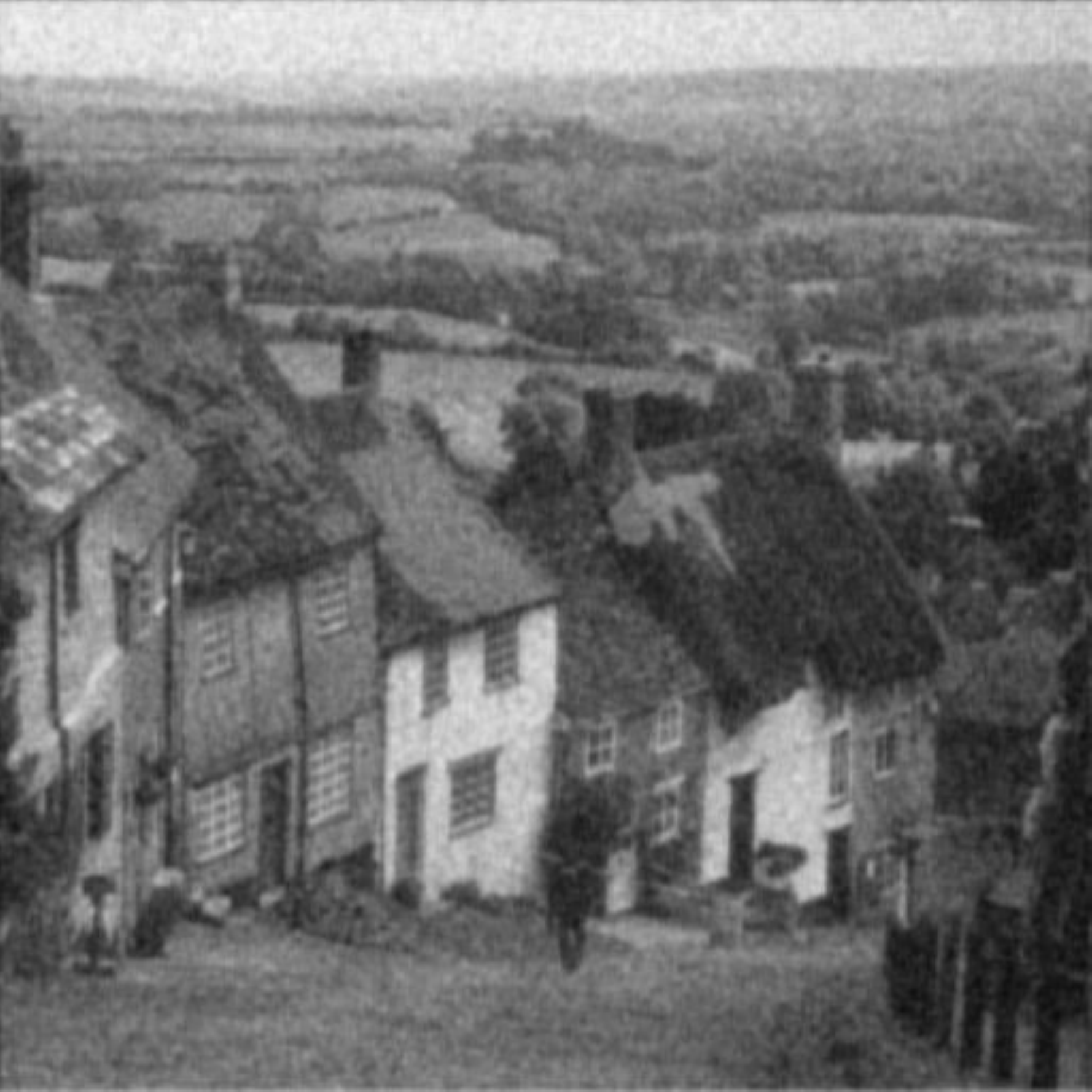}
		\caption{FPGA result.\\PSNR =\\27.825 dB.\\PSNR Gain =\\7.589.}
	\end{subfigure}
	\begin{subfigure}[t]{0.25\columnwidth}\centering
		\includegraphics[width=0.95\columnwidth]{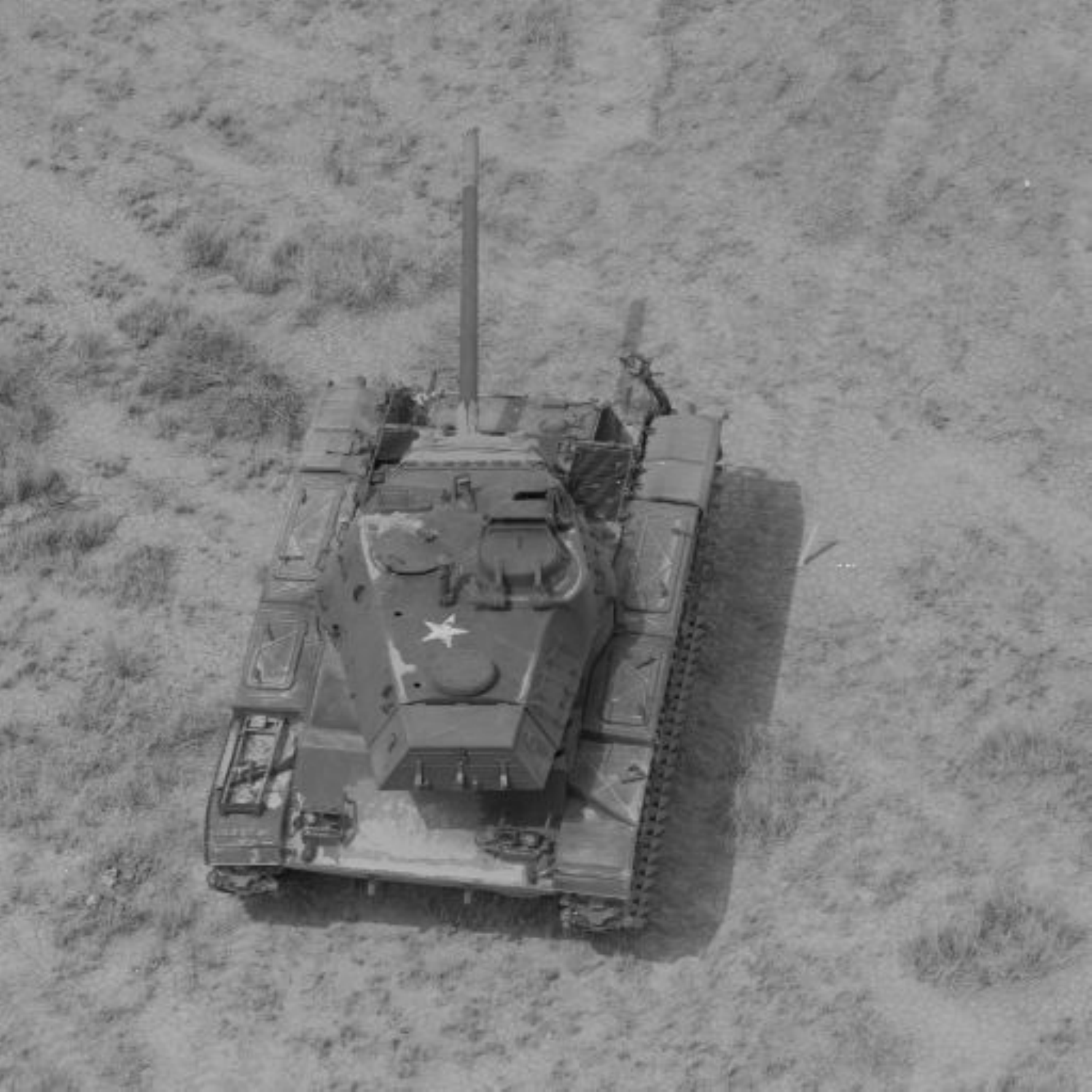}
		\caption{Clean Image.}
	\end{subfigure}%
	\begin{subfigure}[t]{0.25\columnwidth}\centering
		\includegraphics[width=0.95\columnwidth]{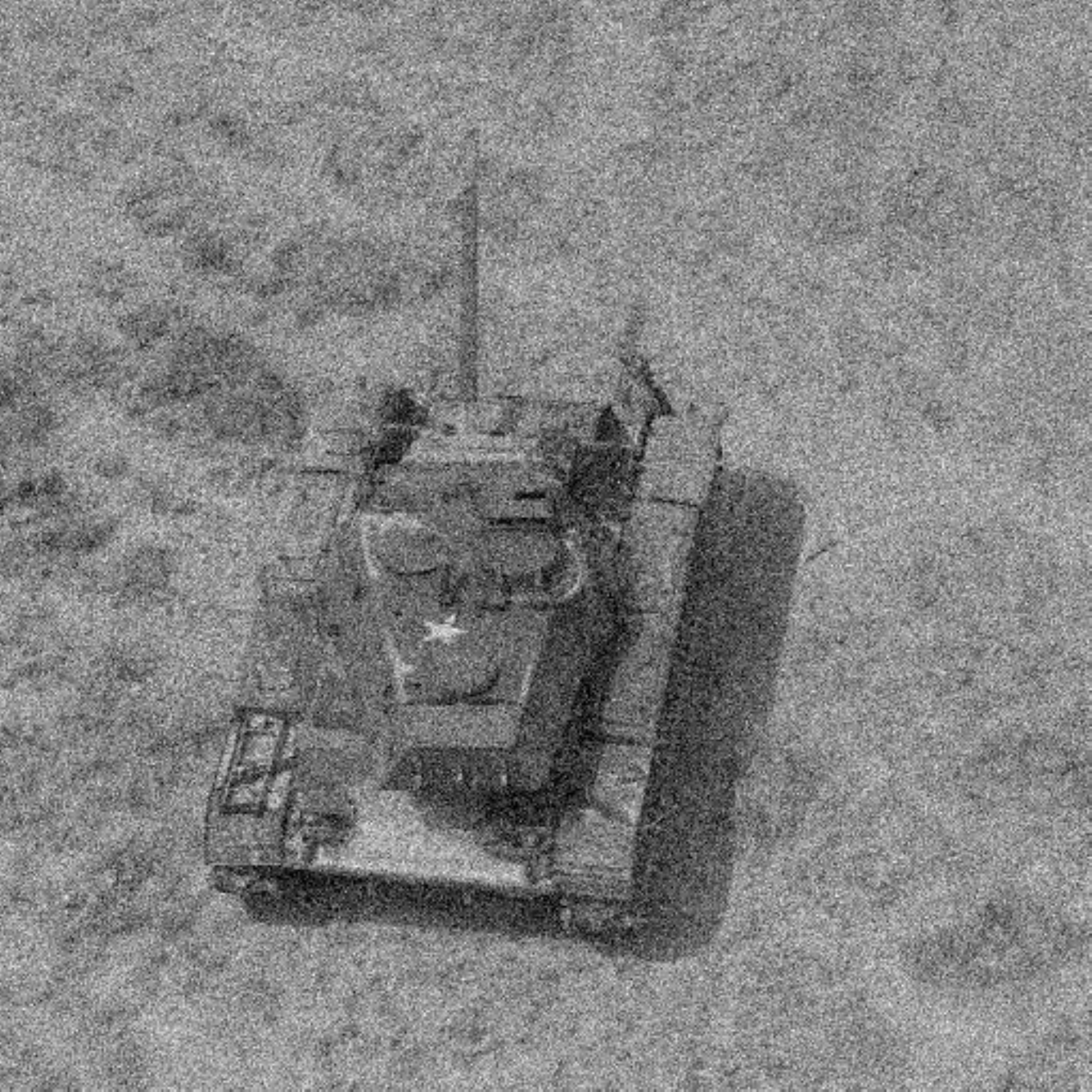}
		\caption{Noisy Image.\\PSNR =\\20.171 dB.}
	\end{subfigure}%
	\begin{subfigure}[t]{0.25\columnwidth}\centering
		\includegraphics[width=0.95\columnwidth]{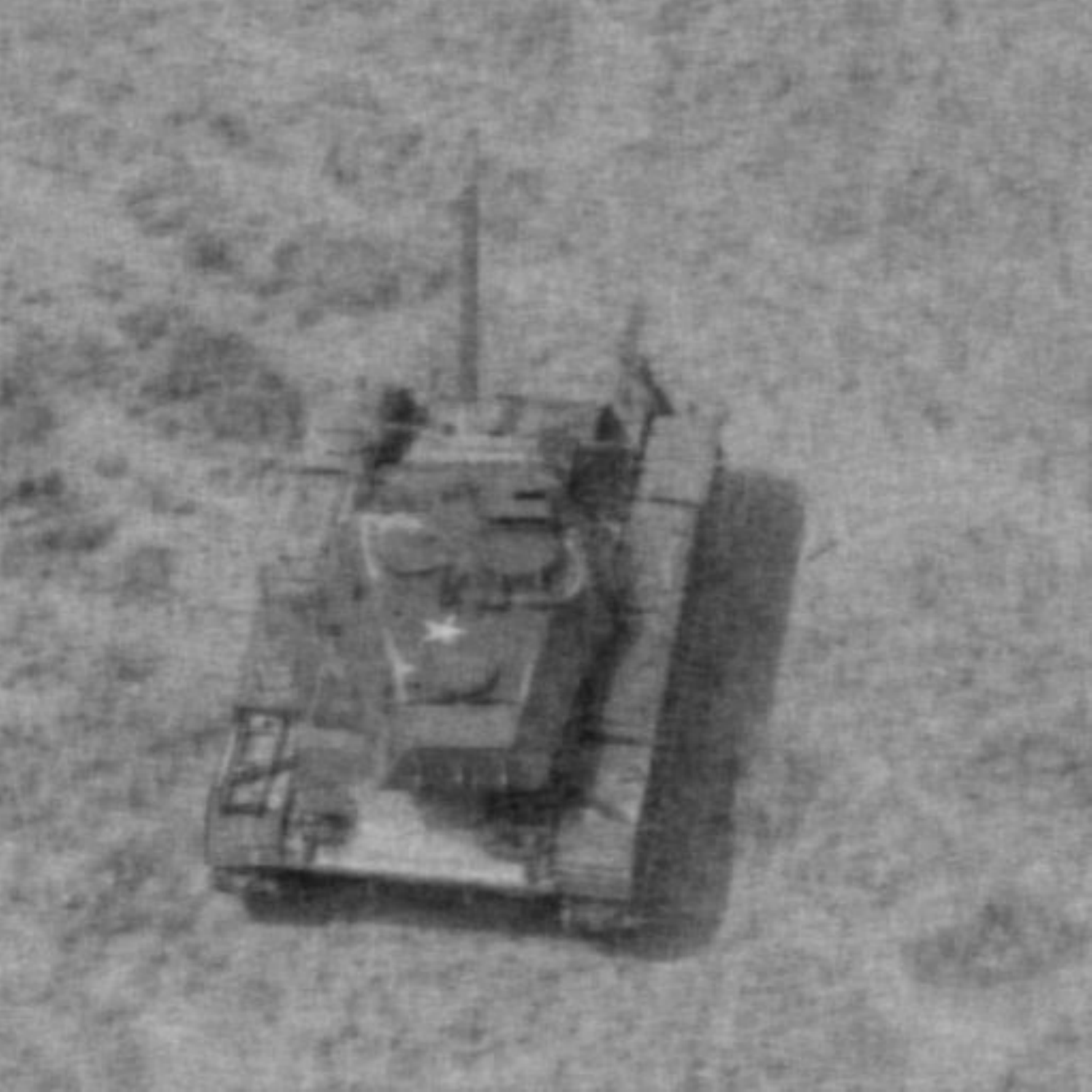}
		\caption{MATLAB\\reference.\\PSNR =\\29.478 dB.\\ PSNR Gain =\\9.307.}
	\end{subfigure}%
	\begin{subfigure}[t]{0.25\columnwidth}\centering
		\includegraphics[width=0.95\columnwidth]{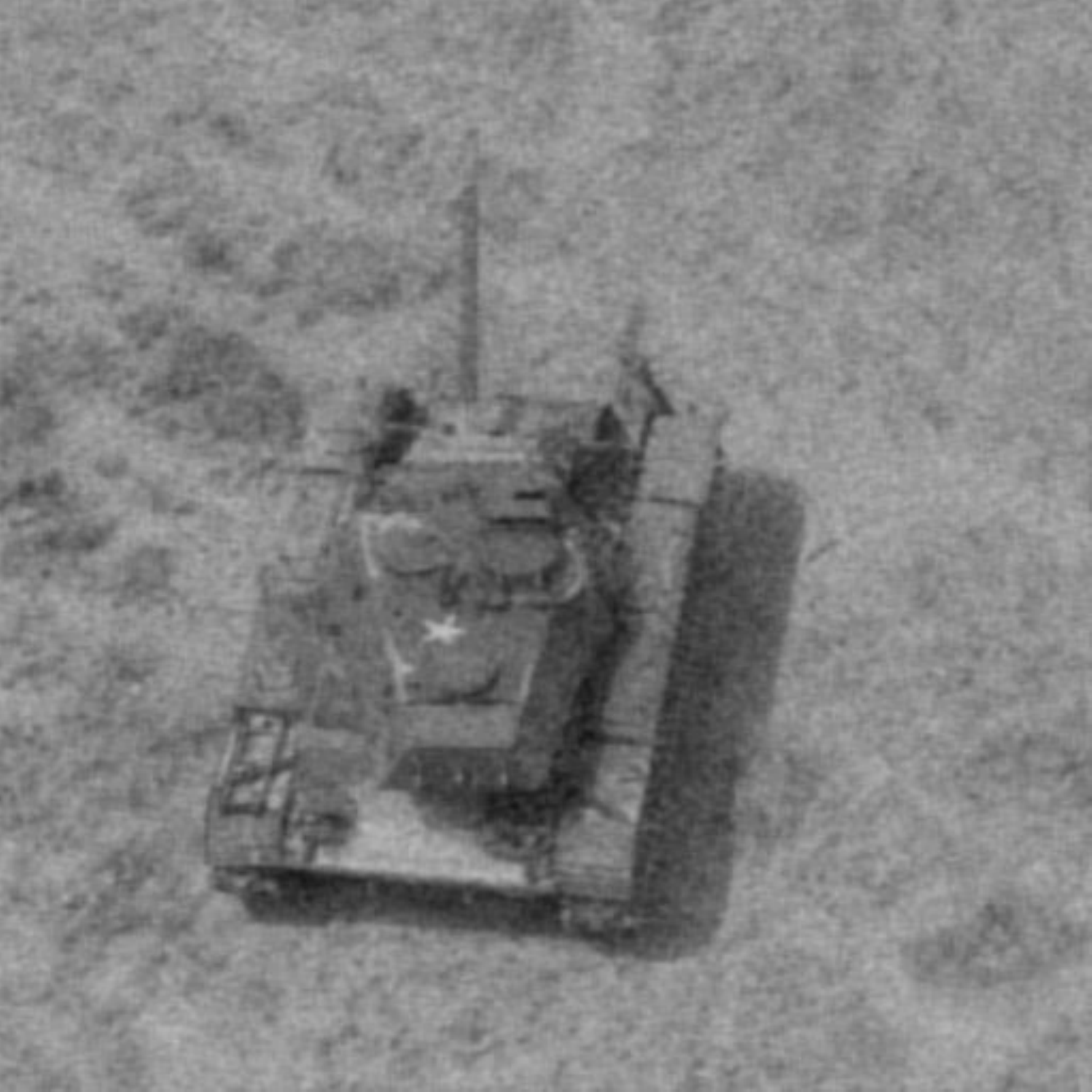}
		\caption{FPGA result.\\PSNR =\\29.274 dB.\\PSNR Gain =\\9.104.}
	\end{subfigure}
	\caption{Comparison of PSNR improvement in MATLAB and FPGA implementations for different test images contaminated with Gaussian noise with $\sigma = 25$. Here MATLAB implementation uses double precision and FPGA uses 16 bit precision.}	
	\label{fig-Comparison_PSNR}
\end{figure}

The proposed architecture is verified using standard test images corrupted with noise and the performance of the design is evaluated using Peak Signal to Noise Ratio (PSNR) which is a commonly used metric defined as
\begin{equation}
\text{PSNR} = 10\log_{10}(\frac{(\max(\yy))^2}{\text{MSE}})\text{dB},
\end{equation} 
$\text{where}\max(\yy) = \underset{i,j}{\max}~y(i,j)$. A comparison of the PSNR obtained from MATLAB and FPGA implementations for images contaminated with gaussian noise with $\sigma = 25$ is shown in Fig.~\ref{fig-Comparison_PSNR}. 
Table \ref{tab-Comparison_Gaussian} -- Table \ref{tab-Comparison_Laplacian} compares the performance for four different images and  three different noise distributions. We observe that the PSNR gain of FPGA implementation is very close to that of MATLAB. 

Another performance metric used is structural similarity (SSIM) index, which measures the similarity between two images. It is defined as
\begin{equation}
\text{SSIM}(I_1,I_2) = \frac{(2\mu_{I_1}\mu_{I_2}+c_1)(2\sigma_{I_1I_2}+c_2)}{(\mu_{I_1}^2+\mu_{I_2}^2+c_1)(\sigma_{I_1}^2+\sigma_{I_2}^2+c_2)},
\end{equation}
where\\
$\mu_{I_1}$ is the average of $I_1$,\\
$\mu_{I_2}$ is the average of $I_2$,\\
$\sigma_{I_1}$ is the variance of $I_1$,\\
$\sigma_{I_2}$ is the variance of $I_2$,\\
$\sigma_{I_1I_2}$ is the covariance of $I_1$ and $I_2$,\\
$k_1 = 0.01$ and $k_2 = 0.03$,\\
$L$ is the dynamic range of the pixel values, and\\
$c_1 = (k_1L)^2,\,\text{and}\, \,c_2 = (k_2L)^2$ are two variables used to stabilise the division with a weak denominator.

Table \ref{tab-Comparison_Gaussian_SSIM}--\ref{tab-Comparison_Laplacian_SSIM} compares the performance of the hardware in terms of SSIM index and Table \ref{tab-Comparison_Exec_Time} gives the execution time across different platforms. It has been observed that the SSIM index we obtain from FPGA and MATLAB are very close and the improvement in execution time is almost hundred times compared to the C implementation.
\section{Conclusion and future work}
We presented an FPGA implementation of the GenRE based image denoising algorithm in~\cite{subramanian2016distribution}. The proposed architecture gives a PSNR gain which is very close to its MATLAB and C counter parts. The improvement in execution time is around hundred times compared to the C implementation. We assumed images of size $512\times512$ for the current implementation and any higher dimensional image can be processed as blocks of $512\times512$. As a future work, we can reduce the block size to $64\times64$ which will reduce the resource utilization and improve the performance. We can also introduce parallelism between  these $64\times64$ blocks which will reduce the execution time. Another improvement will be to extend the architecture for images contaminated with multiplicative noise. 
\begin{table}[h!]
	\centering
	\caption{Comparison of PSNR for Gaussian noise contamination with $\sigma = 25$}
	\label{tab-Comparison_Gaussian}
	\begin{tabular}{llll}
		\hline
		Image     & Input PSNR & PSNR (MATLAB) & PSNR (FPGA) \\ \hline\\
		Lena      & 20.246     & 29.204        & 29.110      \\
		Tank	  & 20.171     & 29.478        & 29.274      \\
		Boat      & 20.268     & 27.621        & 27.512      \\
		House     & 20.236     & 28.194        & 27.825      \\ \hline
	\end{tabular}
\end{table}
\begin{table}[h!]
	\centering
	\caption{Comparison of PSNR for Uniform noise contamination with $\sigma = 25$}
	\label{tab-Comparison_Uniform}
	\begin{tabular}{llll}
		\hline
		Image     & Input PSNR & PSNR (MATLAB) & PSNR (FPGA) \\ \hline\\
		Lena      & 20.172     & 29.212        & 29.114      \\
		Tank	  & 20.168     & 29.453        & 29.264      \\
		Boat      & 20.237     & 27.627        & 27.525      \\
		House     & 20.244     & 28.212        & 27.848      \\ \hline
	\end{tabular}
\end{table}
\begin{table}[h!]
	\centering
	\caption{Comparison of PSNR for Laplacian noise contamination with $\sigma = 25$}
	\label{tab-Comparison_Laplacian}
	\begin{tabular}{llll}
		\hline
		Image     & Input PSNR & PSNR (MATLAB) & PSNR (FPGA) \\ \hline\\
		Lena      & 20.352     & 29.186        & 29.074      \\
		Tank	  & 20.239     & 29.449        & 29.281      \\
		Boat      & 20.377     & 27.631        & 27.532      \\
		House     & 20.407     & 28.167        & 27.769      \\ \hline
	\end{tabular}
\end{table}

\begin{table}[h!]
	\centering
	\caption{Comparison of SSIM index for Gaussian noise contamination with $\sigma = 25$}
	\label{tab-Comparison_Gaussian_SSIM}
	\begin{tabular}{llll}
		\hline
		Image     & Input SSIM & SSIM (MATLAB) & SSIM (FPGA) \\ \hline\\
		Lena      & 0.2740     & 0.7544        & 0.7474     \\
		Tank	  & 0.2618     & 0.6859        & 0.6746      \\
		Boat      & 0.3481     & 0.7072        & 0.7034      \\
		House     & 0.3247     & 0.7214        & 0.7107      \\ \hline
	\end{tabular}
\end{table}
\begin{table}[h!]
	\centering
	\caption{Comparison of SSIM index for Uniform noise contamination with $\sigma = 25$}
	\label{tab-Comparison_Uniform_SSIM}
	\begin{tabular}{llll}
		\hline
		Image     & Input SSIM & SSIM (MATLAB) & SSIM (FPGA) \\ \hline\\
		Lena      & 0.2668     & 0.7530        & 0.7423     \\
		Tank	  & 0.2565     & 0.6866        & 0.6776      \\
		Boat      & 0.3431     & 0.7060        & 0.7034      \\
		House     & 0.3222     & 0.7213        & 0.7105      \\ \hline
	\end{tabular}
\end{table}
\begin{table}[h!]
	\centering
	\caption{Comparison of SSIM index for Laplacian noise contamination with $\sigma = 25$}
	\label{tab-Comparison_Laplacian_SSIM}
	\begin{tabular}{llll}
		\hline
		Image     & Input SSIM & SSIM  (MATLAB) & SSIM (FPGA) \\ \hline\\
		Lena      & 0.2842     & 0.7582        & 0.7495      \\
		Tank	  & 0.2730     & 0.6866        & 0.6791      \\
		Boat      & 0.3614     & 0.7115        & 0.7075      \\
		House     & 0.3403     & 0.7221        & 0.7063      \\ \hline
	\end{tabular}
\end{table}
\begin{table}[h!]
	\centering
	\caption{Comparison of execution times. System configuration and tools used for MATLAB and C implementation: Intel core i5-2300 @
		2.80GHz with 4 GB of RAM, Visual studio 2017 with Intel parallel studio for matrix operations}
	\label{tab-Comparison_Exec_Time}
	\begin{tabular}{ll}
		\hline
		Platform & Execution time (ms) \\ \hline\\
		MATLAB   & 1342                \\
		C        & 380                 \\
		FPGA     & 3.6                 \\ \hline
	\end{tabular}
\end{table}
\bibliographystyle{ieeetr}
\bibliography{MyBibliography}	

\begin{thebibliography}{10}

\bibitem{QSI}
Q.~Corporation, ``Understanding ccd read noise.'' \url{www.qsimaging.com/ccd
  noise.html}, 2008.
\newblock [Online; accessed 19-July-2017].

\bibitem{wilkinson1998digital}
M.~H. Wilkinson and F.~Schut, {\em Digital image analysis of microbes: imaging,
  morphometry, fluorometry and motility techniques and applications}.
\newblock John Wiley \& Sons, 1998.

\bibitem{healey1994radiometric}
G.~E. Healey and R.~Kondepudy, ``Radiometric ccd camera calibration and noise
  estimation,'' {\em IEEE Transactions on Pattern Analysis and Machine
  Intelligence}, vol.~16, no.~3, pp.~267--276, 1994.

\bibitem{gravel2004method}
P.~Gravel, G.~Beaudoin, and J.~A. De~Guise, ``A method for modeling noise in
  medical images,'' {\em IEEE Transactions on medical imaging}, vol.~23,
  no.~10, pp.~1221--1232, 2004.

\bibitem{frost1982model}
V.~S. Frost, J.~A. Stiles, K.~S. Shanmugan, and J.~C. Holtzman, ``A model for
  radar images and its application to adaptive digital filtering of
  multiplicative noise,'' {\em IEEE Transactions on pattern analysis and
  machine intelligence}, no.~2, pp.~157--166, 1982.

\bibitem{SimpleFPGA}
M.~I. AlAli, K.~M. Mhaidat, and I.~A. Aljarrah, ``Implementing image processing
  algorithms in fpga hardware,'' in {\em Applied Electrical Engineering and
  Computing Technologies (AEECT), 2013 IEEE Jordan Conference on}, pp.~1--5,
  IEEE, 2013.

\bibitem{Bilateral1}
A.~Gabiger-Rose, M.~Kube, R.~Weigel, and R.~Rose, ``An fpga-based fully
  synchronized design of a bilateral filter for real-time image denoising,''
  {\em IEEE Transactions on Industrial Electronics}, vol.~61, no.~8,
  pp.~4093--4104, 2014.

\bibitem{Bilateral2}
A.~Gabiger, M.~Kube, and R.~Weigel, ``A synchronous fpga design of a bilateral
  filter for image processing,'' in {\em Industrial Electronics, 2009.
  IECON'09. 35th Annual Conference of IEEE}, pp.~1990--1995, IEEE, 2009.

\bibitem{Bilateralreconfigurable}
S.~D. Dabhade, G.~Rathna, and K.~N. Chaudhury, ``A reconfigurable and scalable
  fpga architecture for bilateral filtering,'' {\em IEEE Transactions on
  Industrial Electronics}, vol.~65, no.~2, pp.~1459--1469, 2018.

\bibitem{hegde2015adaptive}
K.~V. Hegde, V.~Kulkarni, R.~Harshavardhan, and S.~S. David, ``Adaptive
  reconfigurable architecture for image denoising,'' in {\em Parallel and
  Distributed Processing Symposium Workshop (IPDPSW), 2015 IEEE International},
  pp.~196--201, IEEE, 2015.

\bibitem{di2013aidi}
S.~Di~Carlo, P.~Prinetto, D.~Rolfo, and P.~Trotta, ``Aidi: An adaptive image
  denoising fpga-based ip-core for real-time applications,'' in {\em Adaptive
  Hardware and Systems (AHS), 2013 NASA/ESA Conference on}, pp.~99--106, IEEE,
  2013.

\bibitem{katona2005fpga}
M.~Katona, A.~Pi{\v{z}}urica, N.~Tesli{\'c}, V.~Kova{\v{c}}evi{\'c}, and
  W.~Philips, ``Fpga design and implementation of a wavelet-domain video
  denoising system,'' in {\em International Conference on Advanced Concepts for
  Intelligent Vision Systems}, pp.~650--657, Springer, 2005.

\bibitem{subramanian2016distribution}
B.~K. Subramanian, A.~Gupta, and C.~S. Seelamantula, ``A
  distribution-independent risk estimator for image denoising,'' in {\em
  Proceedings of the Tenth Indian Conference on Computer Vision, Graphics and
  Image Processing}, p.~52, ACM, 2016.

\bibitem{Xilinx}
A.~Finnerty and H.~Ratigner, ``Reduce power and cost by converting from
  floating point to fixed point,'' Tech. Rep. WP491 (v1.0), Xilinx, mar 2017.

\bibitem{meyer2000matrix}
C.~D. Meyer, {\em Matrix analysis and applied linear algebra}, vol.~71.
\newblock Siam, 2000.

\bibitem{bubeck2015convex}
S.~Bubeck {\em et~al.}, ``Convex optimization: Algorithms and complexity,''
  {\em Foundations and Trends{\textregistered} in Machine Learning}, vol.~8,
  no.~3-4, pp.~231--357, 2015.

\end{thebibliography}
	
\end{document}